\documentclass[preprint,12pt,]{elsarticle}
\biboptions{numbers,sort&compress}

\usepackage{amssymb}
\usepackage{amsmath}

\usepackage[obeyspaces, spaces, hyphens]{url}
\usepackage{hyperref}
\hypersetup{breaklinks=true}
\usepackage{booktabs}
\usepackage{tabularx}
\usepackage{makecell}
\usepackage{amssymb}
\usepackage{graphicx} 
\usepackage{xcolor}
\usepackage{float}
\usepackage{graphicx}
\usepackage{algorithm}
\usepackage{algpseudocode}
\usepackage{caption}
\usepackage{wrapfig} 
\usepackage{inconsolata}
\usepackage{listings}   
\usepackage[most]{tcolorbox}
\tcbuselibrary{listingsutf8, breakable, skins, theorems}
\usepackage[utf8]{inputenc}
\usepackage[labelfont={bf, normalsize}, textfont=normalsize]{caption}
\usepackage{subcaption}  
\usepackage[frozencache,cachedir=.]{minted}
\usepackage{multirow}
\usepackage{siunitx}
\usepackage{xcolor} 
\DeclareRobustCommand{\textendash}{\unskip\kern.2em\hbox{--}\kern.2em\ignorespaces}

\newtcolorbox[auto counter]{promptbox}[2][]{%
  enhanced,
  float,
  floatplacement=t,
  colback=gray!5!white,
  colframe=gray!50!black,
  sharp corners,
  width=\textwidth,
  boxrule=0.4pt,
  left=4pt,
  right=4pt,
  top=2pt,
  bottom=2pt,
  title={Prompt~\thetcbcounter: #2},
  fonttitle=\bfseries\small,
  before upper={\ttfamily\scriptsize\selectfont}, 
  listing only,
  listing options={
    basicstyle={\scriptsize\ttfamily},      
    language=Java,
    breaklines=true,
    columns=fullflexible,
    showstringspaces=false,
    aboveskip=0pt,
    belowskip=0pt
  },
  #1
}

\lstdefinestyle{jsstyle}{
    language=JavaScript,
    basicstyle=\ttfamily\small,
    keywordstyle=\color{blue},
    commentstyle=\color{green!60!black},
    stringstyle=\color{purple},
    numbers=none,       
    breaklines=true,
    breakatwhitespace=false,
    frame=none,         
    showstringspaces=false,
    tabsize=2
}
\journal{Information and Software Technology}

\begin{document}

\begin{frontmatter}

\title{From Coverage to Causes: Data-Centric Fuzzing for JavaScript Engines}

\author[inst1]{Kishan Kumar Ganguly\corref{cor1}}
\ead{kgangul@ncsu.edu}

\author[inst1]{Tim Menzies}
\ead{tjmenzies@ncsu.edu}

\affiliation[inst1]{organization={Department of Computer Science, North Carolina State University},
             addressline={Raleigh, NC},  
             country={USA}}

\cortext[cor1]{Corresponding author}



\begin{abstract}
\textbf{Context:} Exhaustive fuzzing of modern JavaScript engines is infeasible due to the vast number of program states and execution paths. Coverage-guided fuzzers waste effort on low-risk inputs, often ignoring vulnerability-triggering ones that do not increase coverage. Existing heuristics proposed to mitigate this require expert effort, are brittle, and hard to adapt.\\
\textbf{Objective:} We propose a data-centric, LLM-boosted alternative that learns from historical vulnerabilities to automatically identify minimal static (code) and dynamic (runtime) features for detecting high-risk inputs.\\
\textbf{Method:} Guided by historical V8 bugs, iterative prompting generated 115 static and 49 dynamic features, with the latter requiring only five trace flags, minimizing instrumentation cost. After feature selection, 41 features remained to train an XGBoost model to predict high-risk inputs during fuzzing.\\
\textbf{Results:} Combining static and dynamic features yields over $85\%$ precision and under $1\%$ false alarms. Only 25\% of these features are needed for comparable performance, showing that most of the search space is irrelevant.\\
\textbf{Conclusion:} This work introduces feature-guided fuzzing, an automated data-driven approach that replaces coverage with data-directed inference, guiding fuzzers toward high-risk states for faster, targeted, and reproducible vulnerability discovery. To support open science, all scripts and data are available at \url{https://github.com/KKGanguly/DataCentricFuzzJS}
.

\end{abstract}



\vspace{0.5cm}

\begin{keyword}


Testing, Fuzzing, Static Analysis, Dynamic Analysis,
 JavaScript, V8
\end{keyword}

\end{frontmatter}



\section{Introduction}
\label{intro}
The complexity of modern software has induced a plethora of variability, largely expanding the time and effort required for exhaustive testing. Fuzzing or Fuzz Testing, which is a technique to find  bugs or vulnerabilities, reduces this time and effort by automating generation of inputs in order to exercise unexpected program behavior \cite{li2018fuzzing, chen2018hawkeye}. Coverage-guided fuzzers generate new inputs or mutate existing inputs to reach unexplored program paths. However, its success depends on the assumption that new paths correlate with new vulnerabilities \cite{lemieux2018fairfuzz}. This is inefficient for a complex system like JS engines because a lot of computational resources are spent exploring new but low-risk paths since: 
\begin{itemize}
    \item \textit{It is blind to the high-risk paths} \cite{chen2020savior, li2018fuzzing, wang2024optfuzz}. Since coverage-guidance ignores events that are not tied to new edges or blocks, it fails to prioritize an input that triggers risky JS engine behavior (e.g. JIT bailouts, stressing the garbage collector etc.).
    \item \textit{It destroys high-risk semantically-rich patterns of an input}  \cite{park2020fuzzing, wang2024optfuzz, wang2025patchfuzz}. This is because mutating an input towards exploring new paths can eliminate the potential high-risk behaviors it contains.
\end{itemize}
Later in this paper, we offer experimental results that coverage-driven search fails to retain most of the vulnerability-inducing seeds while repeatedly exploring low-risk regions. To overcome this limitation, fuzzing needs a guidance signal that captures how likely an input is to expose a vulnerability.

In the literature, researchers have addressed the drawbacks of coverage guidance using expert-designed heuristics \cite{wang2024optfuzz, wang2023fuzzjit, park2020fuzzing,wachter2025dumpling, bernhard2022jit, he2021sofi, lee2020montage}. Apart from the manual effort involved, these heuristics often target specific bug classes or are not systematically derived from and validated against a diverse set of historical vulnerabilities. For example, DIE \cite{park2020fuzzing} focuses on preserving "aspects", which they identify as data types and structures, in order to prevent coverage-guided mutations from destroying high-risk behaviors. This was later shown not to generalize to other classes of vulnerabilities \cite{wang2025patchfuzz}, suggesting the existence of additional "aspects". OPTFuzz \cite{wang2024optfuzz} introduced "optimization path coverage," which is effective for finding JIT vulnerabilities but cannot be applied to discover many other types of JS engine bugs. Another line of work aims to reduce the search space by generating more semantically valid inputs \cite{han2019codealchemist, wang2019superion, dinh2021favocado, gross2023fuzzilli, lee2020montage, ye2021automated, xu2024fuzzingneu, wen2023evaluating}. For example, Fuzzilli \cite{gross2023fuzzilli} proposed the FuzzIL Intermediate Representation (IR) to increase semantic correctness. While this line of work improves validity, it can still produce semantically correct but low-risk inputs which leads to inefficient exploration. Our novelty lies in reducing this inefficiency by shifting from manual heuristics to a systematic, data-centric methodology for feature discovery and validation. 

In this work, we leverage the reasoning capabilities of LLMs for automated feature generation. We design a prompting strategy to analyze and generate a diverse set of 115 static and 49 dynamic features (high-risk patterns) from historical vulnerabilities. Applying feature selection via EXtreme Gradient Boosting (XGBoost) model, we show that only 25\% of the combined feature set can predict 160 historical vulnerability-inducing inputs with over $85\%$ precision and $<1\%$ false alarm rate in a time-controlled study. Our automated feature collection strategy not only reduces the total feature search space but also significantly lowers the instrumentation effort, as the generated dynamic features come from just five V8 trace flags the LLM recommended through our prompting pipeline. A time-restricted prudence check at the end shows that a fuzzer guided by this minimal feature set finds real crashes faster. 

The primary challenge to this approach is the vast search space of possible features. Each JS input can be analyzed by countless static code features. Moreover, its execution can generate gigantic trace data. This raises the following questions: how to systematically extract and select the most valuable features from this noise, and how to measure their effectiveness in discovering JS engine vulnerabilities? To answer these, we pose the following questions. 
\begin{enumerate}
    \item[RQ1] How to systematically derive a set of static and dynamic features most indicative of vulnerabilities in JS engine?
    \item[RQ2] How effective is the guidance model in terms of predictive performance?
    \item[RQ3] Do both static and dynamic features contribute to its performance?
    \item[RQ4] What is the minimal feature set beyond which additional features yield diminishing returns?
    \item[RQ5] Can the feature-guided fuzzing process find real bugs?
    \item[RQ6] What is the runtime overhead of collecting these features?
\end{enumerate}

Overall, our contributions are as follows. 
\begin{enumerate}
    \item We present a novel LLM-guided static and dynamic feature generation technique from historical vulnerability-triggering JS inputs.
    \item Our feature importance analysis, which eliminates 75\% of the features without losing predictive performance, shows that \textbf{traditional fuzzers, which treat all code properties as equally important, spend a lot of time mutating irrelevant codes}.
    \item To reduce instrumentation effort, we introduce a new class of dynamic features from LLM-ranked, vulnerability-relevant trace flags which are not widely utilized in prior work. We show that these features work as high-recall detectors and combining these creates a synergistic effect with about $3\times$ gain in precision/recall.
    \item We develop and validate a predictive guidance model using time-aware cross-validation that mimics real-world presentation of bugs over time.
    \item We demonstrate how this predictive model and the features can be utilized in fuzzing.
    \item We release our curated dataset of extracted features and experimentation scripts \footnote{\url{https://github.com/KKGanguly/DataCentricFuzzJS}} to support reproducibility and future research. 
\end{enumerate}

Before proceeding, we digress to clarify a crucial aspect of our methodology. One challenge in comparing different fuzzing strategies
is the time required for each run. This challenge is particularly acute when checking for bugs in well-established tools like JS engines since the more they evolve, the harder it becomes to find new bugs\footnote{For example, we have run standard tools for several weeks of CPU time without yielding any crash in V8, an outcome that is consistent with recent studies showing declining rates of bug discovery in mature, heavily-tested software \cite{wang2024optfuzz, elahi2024forward}}.
Hence, the experiments of this paper are in two parts:
\begin{itemize}
\item A large historical  ``what-if'' analysis where we run data miners over the bugs found after every nine months of JS development. In 2025, data mining can be applied using some heavily optimized algorithms, so this part of our analysis can be practically applied to a very large sample. 
\item A smaller second part where we use the feature found by our data miners to see what new bugs (not in the historical record) can be found. This process is much slower than data mining and so this second part of our study is more limited in scope than the first part.
\end{itemize}
The authority of this paper rests on the first part of this analysis while the second part should be viewed as a prudence check to ensure that the features from the first part are applicable beyond the historical data. 

Note that the computational issues that limited the second part of this analysis is a strong motivator for the first part. Since fuzzing is expensive, it is vital to make full use of all past information when planning future fuzzing. 
 

The paper is organized as follows. Section 2 discusses the terminologies, while Section 3 explains the motivation of this study. Section 4 reviews relevant literature. Section 5 details our methodology for LLM-guided feature extraction, predictive model development, and fuzzer implementation. Section 6 presents our experimental setup and results. We discuss threats to validity and suggest directions for future research in Section 7. Finally, Section 8 concludes the paper with a discussion of our contributions.

\section{Terminology}
Before beginning, we define some terms used in this paper.
\subsection{Fuzzing}
Fuzzing is an automated testing technique that generates and feeds large volumes of test inputs, called seeds, to find bugs or vulnerabilities \cite{zhu2022fuzzing, manes2019art, li2018fuzzing, chen2018hawkeye}. For example, in JS engine, these seeds are JS source codes. This input generation process can be, 1) Black-Box: random or without knowing the system's internal structure as in Radamsa \footnote{https://gitlab.com/akihe/radamsa}, 2) White-box: based on the system's source code as in \cite{godefroid2008grammar, ganesh2009taint} 3) Grey-box: guided by partial knowledge of the system such as coverage, runtime feedback etc. such as AFL \cite{zalewski2014afl}, AFL++ \cite{fioraldi2020afl++}, libFuzzer \cite{libfuzzerLLVM}, Superion \cite{wang2019superion}, Fuzzilli \cite{gross2023fuzzilli}, and many other JS engine fuzzers \cite{park2020fuzzing,wang2023fuzzjit,wang2024optfuzz,xu2024fuzzing, bernhard2022jit, wachter2025dumpling, park2021jest, lima2021exposing, eom2024fuzzing, wang2025patchfuzz, lin2019deity}. Grey-box fuzzers strike a balance between speed and knowledge of the internal structure of the system, making these an ideal choice for fuzzing bulky JS engines.  
\subsection{Fuzzing Strategies}
The effectiveness of fuzzing depends on its strategy to produce inputs. Over the years, mutation-based and generation-based fuzzing strategies have dominated the field. 

\subsubsection{Mutation-Based Fuzzing}
Given a set of initial seeds, known as seed corpus, mutation-based fuzzers apply a series of modifications to create new test cases. These modifications include flipping random bits \cite{zalewski2014afl,fioraldi2020afl++}, AST subtree replacement \cite{wang2017skyfire,wang2019superion,park2020fuzzing}, fragment assembly \cite{lee2020montage,wang2025patchfuzz}, and mutating an intermediate language \cite{gross2023fuzzilli}. Fast execution makes these fuzzers preferable for JS engine fuzzing. However, its effectiveness depends on the quality and diversity of the seed corpus.  

\subsubsection{Generation-Based Fuzzing}
These fuzzers create test inputs from scratch, guided by a predefined model or grammar. This is advantageous because it generates more syntactically valid code \cite{lee2020montage, gross2023fuzzilli}. 
However, since it is not guided by an initial set of seed corpus, it may produce inputs unrepresentative of real-world bugs.

\subsection{Coverage-Guided Fuzzing}
Unguided mutational or generational fuzzers cannot cover all possible program behaviors for a complex system like JS engine, hence, it needs to be guided. 
Coverage guidance remains the dominant technique for JS engine guided fuzzing \cite{gross2023fuzzilli, wang2024optfuzz, park2020fuzzing}. Its central assumption is that exploring new code (or path) leads to the discovery of new bugs. The fuzzer operates as a search algorithm to increase code coverage as a fitness function. 

As many coverage-guided JS engine fuzzers are mutation-based, we discuss their workflow \cite{zhu2022fuzzing, manes2019art} as follows.
\begin{enumerate}
    \item Seed Scheduling: The fuzzer selects an input from its seed corpus, either randomly or weighted by coverage. 
    \item Mutation: The selected seed is mutated to create a new test case.
    \item Execution and Monitoring: The mutated input is fed to an instrumented version of the JS engine. The execution is monitored for crashes or failures, and the code coverage achieved by the input is computed.
    \item Evaluation: If the execution results in a crash, it is saved for later analysis. Otherwise, its coverage data is compared against the total coverage seen so far. 
    \item Corpus Update: If the new input results in new coverage, it is deemed "interesting" and is added to the seed corpus. 
\end{enumerate}

In our work, we argue that this guidance model is inefficient in finding JS engine vulnerabilities. In \S \ref{motivation} and \S \ref{litreview}, we discuss the coverage guided fuzzers for JS engine and their disadvantages. 

\subsection{Feature-Guided Fuzzing}
This is a term coined by this paper. Feature is a specific, measurable characteristic of a program which is used to predict the likelihood of containing a bug or vulnerability. Feature-guided fuzzing is an approach guided by a composite score (e.g. probability from a model) from these features. This guidance method shifts the question from "is this path new" to "does this code look dangerous". We utilize two types of features in this paper. 
\subsubsection{Static Features}
Measurable characteristics of a program which are identified from its source code without needing execution. For example, count of recursive functions, nested exception handling, functional complexity score etc. 
\subsubsection{Dynamic Features}
Dynamic features are measurable properties observable only during the program's execution. These capture runtime events which are usually not discoverable from the source code. For example, turning on V8 trace flags like \texttt{--trace-gc} produces logs of garbage collection events. Dynamic features such as garbage collector call counts can be collected from these logs .  

\section{Motivation}
\label{motivation}
In this section, we explain "Why" we need a new fuzzing approach for JS engine with an example.

The following JavaScript code (Listing \ref{lst:poc_crash}), when run with specific stress-testing flags, triggers a memory corruption bug in the V8 engine. 

\begin{figure}[H]
\captionof{listing}{\normalsize Proof-of-Concept (PoC) code designed to trigger a memory corruption in the V8 JavaScript engine (Chromium Issue 40068612).}
\label{lst:poc_crash}
\small
\centering
\begin{minted}[breaklines, linenos=false]{javascript}
function f0() {
  try {
    for (let v0 = 0; v0 < 100; v0++) {
      const v1 = [0, 0, 1.1];  // Creates JIT complexity
      const v2 = "test";       // Initialize v2 to avoid ReferenceError
      const v3 = eval(Math);   // Creates unusual type states
      try {
        let v4 = v2.substring(v3, v0); // Suppressed error
      } catch (e) {}
      try {
        f0(); // Infinite recursion to trigger JIT & GC
      } catch (e) {}
    }
  } catch (e) {}
}
f0();
\end{minted}
\vspace{-0.4cm}
\end{figure}

The deep recursion of \texttt{f0} calling itself and the GC call forces the JIT to aggressively optimize the function by doing On-Stack Replacement (OSR) \footnote{\protect\url{https://v8.dev/blog/v8-release-79\#osr-caching}}. This is a complex optimization where the engine tries to switch to a faster compiled code in the middle of the function's execution to avoid a stack overflow. However, due to a bug in the stack check mechanism, the engine keeps the wrong type of value in the memory. Running the GC forces the memory to be reorganized causing the type assumption of the JIT to be violated. This triggers the type check, which causes the crash. 

In the following sections, we utilize this example to state our motivation.

\subsection{Coverage Alone is Insufficient for Vulnerability Discovery}
Code coverage is not a direct measure of progress for vulnerability-prone state exploration. In the literature, coverage has been criticized as an ineffcient tool for finding crashes or vulnerabilities \cite{chen2020savior, li2018fuzzing, wang2019superion, bohme2022reliability}, particularly for the JS engine. Using coverage alone is an inadequate and inefficient approach due to the following reasons.   
\begin{itemize}
    \item \textbf{Blindness to data and semantic Logic:} Traditional code coverage only reflects which new codes are run, ignoring the meaning in a program's data and semantics. Hence, a fuzzer that is good at increasing code coverage may not be effective at reaching the meaningful paths that are important to trigger the bugs \cite{wang2019superion, bohme2022reliability}. 
    \item \textbf{Inability to distinguish paths vs. edges:} Traditional edge coverage-guided fuzzers might cover all individual edges over time but never traverse some of the combinations of these edges that lead to bugs \cite{wang2024optfuzz,wang2025patchfuzz}. 

     \item \textbf{State blindness:} As programs are run, they go through a sequence of JS engine internal states or events such as warm-up, optimization, GC events etc., which are not directly observable from code or coverage information. Runtime traces (e.g. via flags in V8) expose these internal states, since developers add them to monitor complex or error-prone areas. Thus, the presence of traces highlight execution paths that developers themselves find risky or crash-prone. Coverage-guided fuzzers overlook these traces and instead rely on seed corpus quality to cover diverse internal states \cite{wang2025patchfuzz}, which is hard to get and verify. 

    \item \textbf{Destructive mutations:} To maximize code coverage, fuzzers aggressively mutate seeds without considering semantic integrity. Hence, highly vulnerable features in seeds are eventually destroyed \cite{park2020fuzzing, wang2025patchfuzz}. 
\end{itemize}
 
In Listing \ref{lst:poc_crash}, the body of the \texttt{f0} function contains a simple loop. A fuzzer may execute a seed with this loop once and all of the internal paths will be covered. Mutations that could satisfy the crash-triggering condition such as deep recursion but do not contribute to new coverage, are not added to the seed corpus. This prevents the corpus to mutate towards the conditions required to trigger the bug. 

We evaluated coverage-guided fuzzing on 160 vulnerability-triggering and $\approx$3000 benign JS code seeds and selected those incrementally based on coverage gain. This effectively simulates the ``interestingness'' discussed earlier assuming a "perfect mutator". Over 10 runs, only $\approx$22\% vulnerable and 1400 benign seeds were selected on average. A large pool of vulnerabilities were ignored because they offered no new coverage. Although some of the selected benign seeds may still hide undiscovered issues, this experiment highlights a core limitation of coverage guidance. It cannot distinguish a benign execution path from a vulnerability triggering one when they both traverse the same control flow regions, and may there miss state or data dependent bugs once the path is superficially covered.

\subsection{Heuristic-Based Semantic Guidance Need Manual Effort}
In the literature, the shortcomings of coverage guidance are often addressed by incorporating manually-engineered rules, observations, or assumptions, which introduce the following limitations.

\begin{itemize}  
    \item \textbf{Lack of adaptivity:} Heuristic rules are fixed at design time and cannot automatically adapt as new types of vulnerabilities emerge.   
    \item \textbf{Limited scope:} Manual heuristics are inherently biased towards specific vulnerability classes. For example, fuzzers designed for JIT bugs \cite{wang2024optfuzz, wang2023fuzzjit} cannot find other classes of vulnerabilities. Even heuristics applicable to all JS engine components often miss bugs that slightly violate the rules. For example, DIE fuzzer's \cite{park2020fuzzing} heuristic of preserving type and structure fail to find bugs requiring control structure alteration \cite{wang2025patchfuzz}.
    \item \textbf{High manual Effort and domain expertise:} In JS engine fuzzing studies \cite{park2020fuzzing, bernhard2022jit, gross2023fuzzilli, he2021sofi, wang2023fuzzjit}, heuristics are often created by human experts. For a complex system like JS engine, which has frequent updates \footnote{Every week for dev builds and every four weeks for stable releases in V8 (\url{https://v8.dev/docs/release-process})} and variations between components in different versions, manually designing and maintaining such heuristics is difficult.
\end{itemize}

Overall, we argue that a lack of a better guidance is one of the reasons for the large runtime and resource requirements reported for finding bugs in JS engines. For example, Fuzzilli found over 17 bugs over the period of six months with 500 CPU cores \cite{gross2023fuzzilli}. Similarly, DIE found 48 new bugs across major JS engines, but required 388-839 CPU cores and distributed processing over a week \cite{park2020fuzzing}. When a fuzzer's search strategy is inefficient at targeting bug-prone states, it must compensate with brute force computation. This provides a strong motivation for us to develop a guidance system that prioritizes these high-risk states. 

\subsection{JS Engine Vulnerabilities Show Recurrent Patterns over Time}
Recently, we collected 220 security bug reports from the Chromium bug repository for V8 (see \S \ref{section:data_collection} for details). An analysis of these historical vulnerabilities showed recurring patterns as follows.  
\begin{enumerate}
    \item Concentration in high-risk Areas: We observe that vulnerabilities are not uniformly distributed. Our analysis shows that over 57\% of crashing inputs contain features explicitly designed to stress the JIT compiler and Garbage Collector, such as internal function calls such as \texttt{\%OptimizeFunctionOnNextCall}, loops with many iterations, explicit \texttt{gc} calls etc. This indicates that exercising these components is more likely to expose vulnerabilities.
    \item Attackers repeatedly target the same vulnerable primitives: The same high-risk code constructs appear across a large fraction of vulnerabilities. For example, $\approx 40\%$ of the vulnerabilities in our dataset utilized typed arrays, which is a common source of memory safety bugs \cite{cve20167288, cve202017053}. This indicates that attackers focus on a small, recurring set of weak points.
    \item Static features have blind spots: Static features often miss some vulnerability patterns. For example, in our dataset, every one out of five PoCs ($\approx 21\%$) with JIT/stress events has no corresponding static indicators.
\end{enumerate}

This motivated us to leverage historical data to identify the recurring features related to JS engine vulnerabilities. We hypothesize that these features are strong predictors of high-risk inputs and can help to find crashes faster. 

\section{Literature Review}
\label{litreview}
\begin{table*}[t!]
    \centering
    \caption{Classification of Related Work in JS Security. Each work is categorized by its primary technical contributions.}
    \label{tab:lit_review}
    \resizebox{\textwidth}{!}{%
    
    \begin{tabular}{c p{3.5cm} p{1cm} c c c c c}
        \toprule
        \textbf{Cites} & \textbf{Title} & \textbf{Year} & \makecell{Coverage \\ Guidance} & \makecell{Heuristics \\ Guidance} & \makecell{Automated Vulnerability \\ Feature Guidance} & \makecell{Static \& Structural \\ Guidance} & \makecell{Dynamic Analysis \\ (Multi-source Traces)} \\
        \midrule
        485 & Skyfire \cite{wang2017skyfire} & 2019 &\checkmark & & & \checkmark & \\ 
        189 & Superion \cite{wang2019superion} & 2019 &\checkmark & & & \checkmark & \\
        189 & CodeAlchemist \cite{han2019codealchemist} & 2019 &\checkmark & & & \checkmark & \\
        140 & DIE \cite{park2020fuzzing} & 2020 & \checkmark & \checkmark & & \checkmark & \\
        128 & Montage \cite{lee2020montage} & 2020 & &\checkmark & & \checkmark & \\
        91 & COMFORT \cite{ye2021automated} & 2020 & & & & \checkmark & \\
        66 & Favocado \cite{dinh2021favocado}  & 2021 & &\checkmark & & \checkmark & \\
        65 & JIT-picking \cite{bernhard2022jit} & 2022 & \checkmark &\checkmark & & & \\
        56 & FUZZILLI \cite{gross2023fuzzilli} & 2023 & \checkmark & \checkmark & & \checkmark & \\
        48 & Sofi \cite{he2021sofi} & 2021 & \checkmark & \checkmark & & \checkmark & \\
        41 & FuzzJIT \cite{wang2023fuzzjit} & 2023 &\checkmark &\checkmark & & & \\
        26 & Jest \cite{park2021jest} & 2021 &\checkmark &\checkmark & &\checkmark & \\
        10 &  CovRL-Fuzz \cite{eom2024fuzzing} & 2024 & \checkmark & & & & \\
        9 & JS-SBST \cite{zhang2023javascript} & 2023 &\checkmark &\checkmark & & \checkmark & \\
        8 & JS Vuln Detect \cite{kluban2024detecting}  & 2024 & & & & \checkmark & \\
        4 & JSFuzz \cite{wen2023evaluating} & 2023 &\checkmark &\checkmark & & \checkmark & \\
        3 & OptFuzz \cite{wang2024optfuzz} & 2024 &\checkmark & \checkmark & & & \\
        2 & PMFuzz \cite{xu2024fuzzingneu} & 2024 &\checkmark & & & \checkmark & \\
        2 & DUMPLING \cite{wachter2025dumpling} & 2025 &\checkmark &\checkmark & & & \\
        - & \textbf{Proposed} & 2025 &\checkmark  & \checkmark & \checkmark & \checkmark & \checkmark \\
        \bottomrule
    \end{tabular}%
    \vspace{-10pt}
    }
    
\end{table*}

\begin{wrapfigure}[13]{l}{0.55\textwidth}
\vspace{-0.4cm}
\centering
\includegraphics[width=0.9\linewidth]{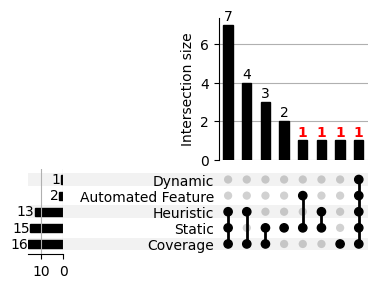}
\caption{UpSet plot of five research categories in JS engine fuzzing. Our paper is the right-most entry}
\label{fig:venn_diagram}
\end{wrapfigure}

\vspace{-5pt}
To contextualize this work, we conducted a comprehensive survey of research on guided fuzzing for JavaScript engines. We searched the combinations of four terms, which are "JavaScript engine fuzzing", "fuzzing guidance", "dynamic analysis", and "automated feature discovery" across scholarly databases covering publications from 2019 to 2025.
We extracted the top 1000 results sorted by citation counts, prioritizing papers published in top-tier conferences and journals in software security, programming languages, and automated testing. To ensure relevance and impact, we filtered out all papers with less than 8 citations from before 2024 and included all papers from 2024–2025 with at least one citation. This process yielded 37 papers, which we reviewed and identified 18 publications with direct contributions in any of the core categories of our study (see Table \ref{tab:lit_review}). One more paper was added through snowballing. The resulting set is summarized in Table \ref{tab:lit_review} and the UpSet plot in Figure \ref{fig:venn_diagram}.

Our analysis reveals a research gap. Although existing works focus on subsets of the five categories, none addresses all (see the intersection of all categories in Figure \ref{fig:venn_diagram}). Our work fills this research gap by systematically integrating all five. Moreover, no prior work besides ours employs both automated vulnerability feature guidance and dynamic analysis that leverages multiple trace sources. 
\vspace{0.4cm}
\subsection{Coverage Guidance}
Coverage-guidance is the foundational paradigm for state-of-the-art grey-box fuzzers  \cite{afl, fioraldi2020afl++, libfuzzerLLVM, honggfuzz}. 15 out of 19 fuzzers we studied use coverage guidance (Table \ref{tab:lit_review}). Approaches such as CovRL-Fuzz are purely coverage-guided where coverage is the primary reward signal used to train its LLM-based mutator using reinforcement learning \cite{eom2024fuzzing}. 

However, as we discuss in our motivation, relying on coverage alone is inefficient. This is evident as many fuzzers couple coverage feedback with other techniques. For example, Fuzzilli uses an IR and defines several mutations over it on top of coverage guidance \cite{gross2023fuzzilli}. DIE enhances coverage guidance with "aspect-preserving mutation" to preserve the semantic quality of the seeds \cite{park2020fuzzing}. JIT-Picking and FuzzJIT combine it with heuristic-based differential testing oracles \cite{bernhard2022jit, wang2023fuzzjit}. OptFuzz introduces an optimization path coverage metric for JIT bugs \cite{wang2024optfuzz}. These hybrid fuzzing approaches show that coverage guidance benefits significantly from complementary techniques. 

\subsection{Heuristics-Driven and Static \& Structural Guidance}
To improve on basic coverage, state-of-the-art fuzzers integrate the understanding of program structure and semantics. We classify these works into two categories which are heuristics-driven and static \& structural guidance. 

Heuristic-driven fuzzers use manually-crafted rules and often target specific bug classes. For example, heuristics are widely utilized in finding JIT compiler bugs \cite{wang2023fuzzjit, gross2023fuzzilli, bernhard2022jit, wachter2025dumpling, wang2024optfuzz}. FuzzJIT uses specialized templates to trigger JIT optimization \cite{wang2023fuzzjit}. Fuzzilli manually constructs a custom IR and mutation strategies \cite{gross2023fuzzilli}. JIT-Picking, a differential fuzzer for JIT, custom probing mechanism and rule-based removal of non-deterministic code constructs to support differential oracle \cite{bernhard2022jit}. Dumpling, another JIT differential fuzzer, uses manually designed instrumentation points to extract fine-grained execution state for constructing differential oracles \cite{wachter2025dumpling}. OptFuzz employs the heuristic that paths with complex preconditions are more to trigger JIT vulnerabilities and proposes a new coverage metric called "Optimization Trunk Path" \cite{wang2024optfuzz}. Manual rules also appear in fuzzers that broadly target the JS engine. For example, DIE heuristically preserves type and structure of seeds \cite{park2020fuzzing}. SoFi leverages the heuristic that type information improves semantic validity to guide manual semantic strategies and uses handcrafted rules to repair generated test cases \cite{he2021sofi}. Montage utilizes an observation that most buggy fragments come from regression tests to generate seeds \cite{lee2020montage}. While these heuristics help to find vulnerabilities, they are limited by the manual effort required, narrow scope, and lack of adaptivity as discussed in \S \ref{motivation}.

The second category of work uses static \& structural guidance, which focuses on generating syntactically/semantically valid inputs through automated analysis. Grammar-aware tools that use ASTs, constraints, API specifications, or intermediate representation to guide generation belong to this category, such as CodeAlchemist \cite{han2019codealchemist}, Superion \cite{wang2019superion}, Favocado \cite{dinh2021favocado} and Fuzzilli \cite{gross2023fuzzilli}. This category also includes neural models proposed by Montage \cite{lee2020montage}, COMFORT \cite{ye2021automated}, and PMFuzz \cite{xu2024fuzzingneu} aimed at producing valid code. A recent work by Wen et al. demonstrates that prioritizing valid input seeds with code structure (a.k.a. code static features) is highly effective \cite{wen2023evaluating}. This finding, along with our earlier observations (see \S \ref{motivation}), motivated using static and dynamic features to complement approaches that generate valid inputs.

\subsection{Emerging Trends: Vulnerability and Behavior-Driven Guidance}
State-of-the-art JS engine fuzzers are limited in utilizing past exploits. Although these are often used as seed corpus, fuzzers utilizing coverage or static \& structural guidance prioritize exploration and correctness instead of failure patterns. The idea of vulnerability feature guidance models is to shift the focus from modeling correctness to modeling failure. Prior work in this area is limited to manual heuristic extraction and seed corpus construction using past exploits \cite{lee2020montage, wang2025patchfuzz}. However, the automated discovery of features from historical exploits to directly model failure remains largely unexplored.

The second emerging trend is to incorporate dynamic signals. Dumpling, for example, instruments the JS engine to collect predefined execution states for differential fuzzing \cite{wachter2025dumpling}. JS engines like V8 expose hundreds of tracing flags to log different runtime behaviors. A plethora of trace information with automated feature discovery can help prioritize the dynamic signals relevant to historical crashes,instead of using predefined dynamic signals. This direction of leveraging multi-source trace data from engine runtime flags is yet to be explored.

\begin{figure}[t!] 
    \centering
    \includegraphics[width=\textwidth]{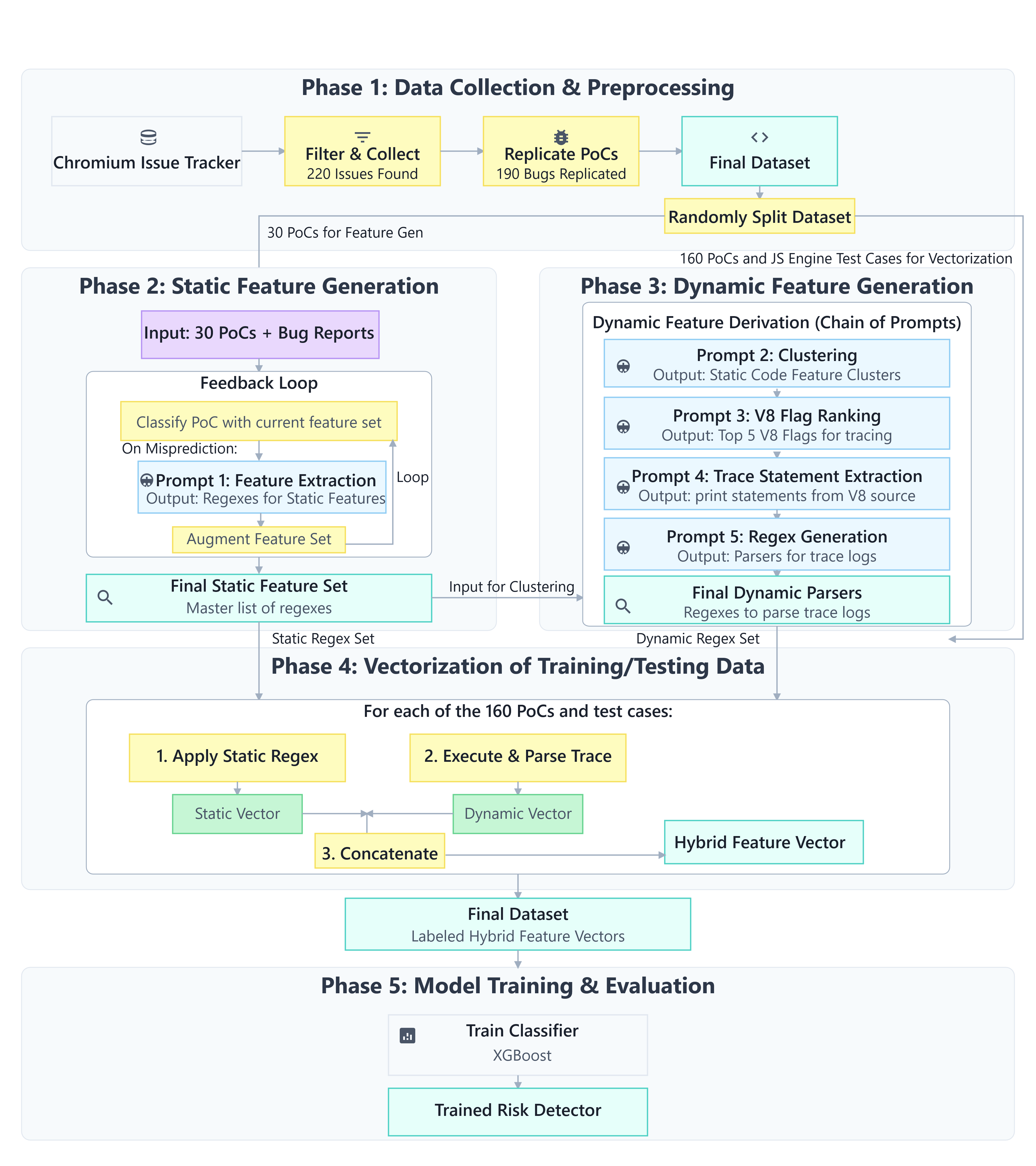}
    \caption{Overview of the Proposed Method}
    \label{fig:overview}
\end{figure}

\section{Methods}
This section describes our methodology including the dataset description, LLM-based feature extraction process, and the construction of the prediction model. It answers the following research question. 

\textbf{RQ1: How to systematically derive a set of static and dynamic features most indicative of vulnerabilities in JS engine?}

Furthermore, we discuss how the feature set and the trained model are applied to fuzzing, which forms the basis for the prudence check. 

\subsection{Overview}
Our method, shown in Figure \ref{fig:overview}, uses a Large Language Model (LLM) to automate feature engineering for a vulnerability prediction model.

First, we construct a dataset of 190 V8 vulnerability-triggering PoCs collected from the Chromium bug repository \footnote{https://issues.chromium.org/issues} and non-vulnerable test cases. Using 30 randomly sampled PoCs from this set, we prompted an LLM to iteratively generate and refine a set of static code features in the form of regular expressions that accurately predicts the given high-risk PoCs. Subsequently, the LLM leveraged these static features and runtime flag documentation to derive the five most relevant runtime flags. Finally, given the V8 source code for these flags, the LLM extracted the trace information schema from which it generated relevant dynamic features in regex format. 

We used these combined features to train an XGBoost machine learning model using the remaining 160 PoCs. This model was used to predict the vulnerability likelihood of a seed.

\subsection{Phase 1: Data Collection \& Preprocessing}
\label{section:data_collection}
The process for collecting the positive (vulnerable) and negative samples are given below. 
\subsubsection{Positive Samples (Vulnerable PoCs)}
We queried the Chromium bug repository with a search query \footnote{\url{https://issues.chromium.org/issues?q=componentid:(1456800|1456566|1456959|1456177|1456567|1456490) status:(fixed|verified) type:vulnerability created:2019-01-01..2025-06-30}} which retrieves fixed or verified bug reports related to the vulnerabilities within the V8 runtime. Filtering for bug reports with available reproduction PoCs and issue description, we obtained 220 reports. 

After this, we ran all of the PoCs through an automated reproduction process as we explain below. This reproduction process is necessary to collect crash-relevant traces for dynamic feature extraction. Since building many versions of V8 takes substantial time, we prebuilt binaries provided by the JavaScript (engine) Version Updater (jsvu) \footnote{\url{https://github.com/GoogleChromeLabs/jsvu}} tool. 

We enumerated all the V8 versions every first and last day of the month starting from January, 2019 to June, 2025. Given a PoC, we extracted its issue date and the crash-triggering runtime flags from the issue text, then ran it through the JS engine versions up to one year prior to the issue date with the flags enabled. This time window was selected to complete the experiment in a reasonable time. For each run, we collected all the generated trace data and looked for keywords such as Crash, Segmentation fault, OOB, etc. to label those as positive. 195 PoCs were reproduced by the automated process. After this, we manually validated all the 195 PoCs' execution to find false positives. 5 PoCs were discarded as false positive, which occurred due to error messages from unsupported flags in the corresponding version. The final positive dataset consisted of 190 PoCs, out of which we use 30 to identify static code features using LLM. 

\subsubsection{Negative Samples}
For this study, the negative samples were collected from the official V8 test suites. Specifically, for each time-based fold of the experiment, the test cases from the mjsunit suite included with the last known stable build of V8 in that period were used as the negative set. We filter the test files to exclude all that contains keywords such as regress, bug, crash etc. in order to avoid labeling a past vulnerability as negative and to avoid overlap with the collected PoCs. 

These tests contain both naive and curated, often adversarial test cases designed to test edge-case behaviors of the engine, ensure specification compliance, and prevent regressions. This choice implies that our classifier is trained to distinguish "crashing complex code" from "non-crashing complex code", rather than simply detecting code complexity.

\begin{promptbox}[label={prompt:structure}]{Structured Prompt Template}

\textbf{1. Identification:} \\
\textit{[Role, expertise, or actor description]}

\vspace{3pt}

\textbf{2. Task Description:} \\
\textit{[High-level task or objective]}

\vspace{3pt}

\textbf{3. Constraints and Requirements:} \\
\textit{[Restrictions, requirements, or conditions]}

\vspace{3pt}

\textbf{4. Inputs:} \\
\textit{[Data, code, or information provided]}

\vspace{3pt}

\textbf{5. Expected Output Format:} \\
\textit{[Format and structure of the expected response]}

\end{promptbox}
\begin{promptbox}[label={prompt:static}]{Static Code Feature Generation Prompt}

\textbf{1. Identification:}

You are an expert cybersecurity analyst specializing in JavaScript engine vulnerabilities.

\vspace{3pt}

\textbf{2. Task Description:}

Design a set of general, measurable, static code features to detect the class of vulnerability demonstrated in the provided Proof-of-Concept (PoC).

\vspace{3pt}

\textbf{3. Constraints and Requirements:}

Crucially, your proposed features must be general and designed to avoid overfitting to this specific PoC. They should identify the underlying vulnerability pattern, not just the superficial implementation details of this one example.

For each feature, you must also provide a regular expression to extract it.

\vspace{3pt}

\textbf{4. Inputs:}

PoC Source Code:

\smallskip

\noindent
[PoC JavaScript code here]

\vspace{3pt}

Issue Description:

\smallskip

\noindent
[Issue text, e.g., from the bug report]

\vspace{3pt}

\textbf{5. Expected Output Format:}

Respond with ONLY a JSON object. Each feature in the list must have a name, description, a functional regex, and a rationale explaining how it generalizes.

\end{promptbox}

{\setlength{\textfloatsep}{6pt}
\setlength{\intextsep}{6pt}
\begin{algorithm}[ht]
\captionsetup{skip=2pt}
\caption{Iterative LLM-driven Static Feature Extraction}
\label{alg:llm_feature_extraction}
\scriptsize
\begin{algorithmic}[1]
\Require PoC set $P$ with bug reports, max iterations $I$, demonstration count $n$
\Ensure Curated static feature set $F$

\State Initialize $F \gets \emptyset$
\State Select subset $S \subset P$, $|S|=n$
\For{$i = 1$ to $I$}
    \For{each $p$ in $S$}
        \While{$p$ is misclassified}
            \State Prompt LLM with bug report, source code, and $F$
            \State Extract features $F_p$ and update $F \gets F \cup F_p$
        \EndWhile
    \EndFor
    \State Curate $F$ to remove overfitting features
    \State Select next PoC $p_{test} \in P \setminus S$
    \State Apply $F$ to $p_{test}$ and prompt LLM \emph{without source code} for classification
    \If{$p_{test}$ misclassified}
        \State Add $p_{test}$ to $S$
    \Else
        \State Remove $p_{test}$ from candidate pool
    \EndIf
\EndFor
\State \Return $F$
\end{algorithmic}
\end{algorithm}

}\noindent
\subsection{Phase 2 and 3: Feature Generation}
In Phase 2 and 3, we use GPT 4.1 \footnote{\url{https://openai.com/index/gpt-4-1}}, the latest model available at the time of this study, to extract features. Inspired by the effectiveness of LLMs in feature discovery shown in recent studies \cite{nam2024optimized, zhang2024dynamic, abhyankar2025llm}, we propose an LLM-driven feature discovery approach using historical vulnerabilities. However, our method differs from the prior work in the following ways.
\begin{enumerate}
    \item Our primary inputs are PoC source codes (in Phase 2) and their static features (in Phase 3), not tabular dataset descriptions as in recent studies \cite{nam2024optimized, abhyankar2025llm}. We augmented these inputs with contextual information to improve task performance. For example, in Phase 2, bug report along with the corresponding PoC source code, and in Phase 3, the V8 flag documentation and tracing source codes were included as context. A challenge was to keep the contextual information, particularly the tracing source codes, under token limit of LLMs. Our solution was to filter out the non-tracing functions. Additionally, unlike prior studies \cite{nam2024optimized, abhyankar2025llm}, we chose to extract features from the historical PoCs  rather than asking the LLM to generate them from scratch because this method produced a more verifiable feature set. 
    \item Methods in the literature often generate features from task descriptions in a single pass. In contrast, our method utilizes a multi-step iterative feedback loop. After generating features from one PoC, we evaluated these by applying to a different PoC and prompted the LLM to refine or add features until the PoC was correctly classified. 
    \item Fully automated feature generation methods can produce features that are too overfitting to a single case. Our method includes a human curation step after feature generation, where we removed overfitting (e.g. exact variable names, constants, etc.) or irrelevant features (e.g. nonexistent trace data features), and re-evaluated the PoCs using the LLM until correct classification was obtained. 
\end{enumerate}

Our prompt structure is shown in Prompt \ref{prompt:structure}. This consists of five components, which are identification, task description, constraints, inputs, and expected output format. Identification specifies the role of the LLM, in our case, an expert cybersecurity analyst specializing in JavaScript engine vulnerabilities. Lu et al., showed that identification significantly improves LLM's performance \cite{lu2024grace}. The prompt then states the task to perform, lists all constrains and requirements, and specifies the inputs and the desired output format. We experimented with the prompt structure removing different components such as the identity, constraints, input format, and output format. Overall, prompts formatted in this structure consistently performed better in terms of the number of overfitting features and usable output.

The following sections describe our approach to static feature generation (\S \ref{staticfeat}) and dynamic feature generation (\S \ref{dynamicfeat}).
\vspace{-0.3cm}
\subsubsection{Phase 2: Static Feature Generation}
\label{staticfeat}
Algorithm \ref{alg:llm_feature_extraction} shows the iterative LLM-driven static feature extraction algorithm. The inputs are all the PoCs with bug reports, max iterations, and demonstration count which is the number of PoCs run without classification feedback. We used a max iteration of 100 and demonstration count of 1. Notably, All our PoCs were correctly classified within this limit.

For each PoC, using  Prompt \ref{prompt:static}, We provided the LLM with the bug report, PoC source code and the current feature set to generate features until it is correctly classified as a vulnerability triggering PoC (Line 4-7). Next, we manually curated the feature set $F$ to remove overfitting features (Line 10). We then selected the next unprocessed PoC, applied $F$, and provided the resulting feature values to the LLM (Line 11-12). The PoC source code itself was not provided at this stage because the LLM may have the corresponding PoC information in its training data, which could cause data leakage. If the PoC was misclassified, we added it back to the candidate pool for feature generation until it was correctly classified (Line 13-16). Following this process, we collected 115 static code features.

\begin{promptbox}[label={prompt:cluster}]{JavaScript Static Feature Clustering Prompt}

\textbf{1. Identification:}

You are an expert cybersecurity analyst specializing in JavaScript engine vulnerabilities.
\vspace{3pt}

\textbf{2. Task Description:}

You have a list of 114 static code features derived from analyzing JavaScript Proof-of-Concept (PoC) programs that lead to crashes. Your task is to cluster these features into semantically related groups. 
\vspace{3pt}

\textbf{3. Constraints and Requirements:}

Base grouping on semantic similarity, not superficial wording. Clusters must be concise, mutually exclusive, and non-overlapping. Generate the minimum number of clusters that still preserve distinct meanings.

\vspace{3pt}

\textbf{4. Inputs:}

\noindent
[list of 114 code static features here]

\vspace{3pt}

\textbf{5. Expected Output Format:}

Respond with ONLY a JSON object containing an array of clusters. Each cluster must have: a cluster name and the list of features in that cluster.

\end{promptbox}
\begin{promptbox}[label={prompt:flagselect}]{Top V8 Flags for Crash Prediction Prompt}

\textbf{1. Identification:}
You are an expert cybersecurity analyst specializing in JavaScript engine vulnerabilities.

\vspace{3pt}

\textbf{2. Task Description:}

You have a set of code static feature clusters and the help text from various JavaScript V8 engine flags. Your task is to identify the ten (10) most important flags for crash prediction.

\vspace{3pt}

\textbf{3. Steps to Perform:}

- Analyze the provided feature clusters to understand the primary categories of code properties associated with crashes (e.g., memory management, type confusion, object layout).
\vspace{3pt}

-Review the descriptions of each V8 flag to understand its purpose. 
\vspace{3pt}

-Cross-reference the two sets of information. The most important flags will be those that enable tracing or modify behavior directly related to the activities described by the feature clusters. 

\vspace{3pt}
-Report the top 10 flags by name, ordered from most to least important.

\vspace{3pt}

\textbf{4. Inputs:}

Feature Clusters from Prompt 2:

[Paste the clustered output from Prompt 2 here]

\vspace{3pt}

V8 Flag Help Text Descriptions:

[Paste the V8 flag help text here]

\vspace{3pt}

\textbf{5. Expected Output Format:}

Respond with ONLY a JSON array of objects. Each object must have the V8 flag name and a short explanation of why it is important for crash prediction.
\end{promptbox}
\vspace{-0.5cm}

\begin{promptbox}[label={prompt:flagprint}]{Identify Print Statements for V8 Flag Tracing}

\textbf{1. Identification:}

You are an expert cybersecurity analyst specializing in JavaScript engine vulnerabilities.
\vspace{3pt}

\textbf{2. Task Description:}

You need to identify the \texttt{print} statements within the V8 source code that emit trace data when the flag \texttt{--[FLAG\_NAME]} is enabled. Using all the provided information, locate and extract the full \texttt{print} statements (including format strings and arguments) that are executed when \texttt{--[FLAG\_NAME]} is active.

\vspace{3pt}

\textbf{3. Constraints and Requirements:}

Include only statements that directly emit trace or logging output when the flag is active. Ensure the list is exhaustive.

\vspace{3pt}

\textbf{4. Inputs:}

Flag Name: \texttt{--[FLAG\_NAME]}

\vspace{3pt}

V8 Documentation for \texttt{--[FLAG\_NAME]}: [Add the relevant V8 documentation]

\vspace{3pt}

V8 Source Code for \texttt{--[FLAG\_NAME]}: [Add the relevant V8 source code files where this flag is implemented/used]

\vspace{3pt}

\textbf{5. Expected Output Format:}

Provide ONLY an exhaustive list of all matching \texttt{print} statements, including the full format string and arguments.

\end{promptbox}
\vspace{0.5cm}
\subsubsection{Phase 3: Dynamic Feature Generation}
\label{dynamicfeat}
In this phase, we extracted dynamic features using the LLM to capture signals from trace data relevant to triggering vulnerabilities. The whole process is shown in Algorithm \ref{alg:llm_dynamic_feature_extraction}. We used four prompts for this purpose. The first prompt (Prompt \ref{prompt:cluster}) asked the LLM to cluster the 114 static code features into semantically related groups. Clustering was done because we observed that without it, LLM recommended flags relevant to only certain features while ignoring others, and also suggested irrelevant flags. For example, without clustering, it often missed flags relevant to object layout and inline caches, such as \texttt{--log-maps} and \texttt{--trace-ic} which enabled extraction of dynamic features with high feature importance. 
\begin{promptbox}[label={prompt:regexgen}]{Regex Generation for V8 Trace Print Statements}

\textbf{1. Identification:}

You are an expert cybersecurity analyst specializing in JavaScript engine vulnerabilities.

\vspace{3pt}

\textbf{2. Task Description:}

You have a collection of \texttt{print} statements extracted from the V8 source code. These statements log trace data used to identify JavaScript engine vulnerabilities. Generate a list of regular expressions that can parse the output generated by these \texttt{print} statements.  

\vspace{3pt}

\textbf{3. Constraints and Requirements:}
Each regex should robustly match the pattern and extract variable data (e.g., pointers, integers, strings) into capturing groups. Analyze the structure, constant strings, and format specifiers (such as \texttt{\%p}, \texttt{\%d}, \texttt{\%s}, etc.) in the provided statements to create regex patterns. The goal is to create a set of parsers for vulnerability prediction in V8.

\vspace{3pt}

\textbf{4. Inputs:}

\smallskip

\noindent
[The complete list of all \texttt{print} statements extracted from all runs of Prompt \ref{prompt:flagprint}]

\vspace{3pt}

\textbf{5. Expected Output Format:}

Provide ONLY a JSON array where each element corresponds to one \texttt{print} statement and contains a regular expression string that matches the print statement output and a brief explanation of what the regex captures.
\end{promptbox}

\begin{algorithm}[t!]
\caption{Phase 2: Dynamic Feature Generation}
\label{alg:llm_dynamic_feature_extraction}
\scriptsize
\begin{algorithmic}[1]
\Require Static features $F_{static}$, flag documentation, expert-ranked flags $E$
\Ensure Curated dynamic features $F_{dynamic}$

\State Cluster static features into groups $C$ using LLM (Prompt~\ref{prompt:cluster})

\State Initialize recommended flags set $R \gets \emptyset$

\ForAll{cluster $c \in C$}
    \State Query LLM (Prompt~\ref{prompt:flagselect}) to get top 10 flags $R_c$ for cluster $c$
    \State Update $R \gets R \cup R_c$
\EndFor

\State Retrieve expert ranks for flags in $R$ from $E$

\State Sort $R$ by expert ranks; break ties using LLM ranking

\State Select top 5 flags $S$ from sorted $R$

\State Initialize print statements set $P \gets \emptyset$

\ForAll{flag $f \in S$}
    \State Obtain V8 logging source code for flag $f$
    \State Extract trace-printing statements using LLM (Prompt~\ref{prompt:flagprint}), add to $P$
\EndFor

\State Infer trace schema from print statements $P$

\State Use LLM to generate candidate dynamic features based on trace schema

\State Manually validate and adjust candidate features

\State \Return final curated dynamic feature set $F_{dynamic}$
\end{algorithmic}
\end{algorithm}

Using these feature clusters and V8 flag documentation (help text), we used Prompt \ref{prompt:flagselect} to get top 10 recommended flags representing all of the feature clusters. Then, these flags were curated and reduced to one flag per cluster. The reason we initially did not extract five flags is that LLMs often fail to follow structural constraints \cite{banerjee2025crane, roh2025break}. Hence, we slightly relaxed the constraint and then curated the list. The curation was done as follows. 
\begin{enumerate}
    \item We obtained a larger list of flags ranked according to criteria provided by experienced industry security experts. These flags were ranked in 0-16 scale where higher means more relevant. 
    \item We obtained the expert rankings for the LLM-ranked flags.
    \item We picked the top 5 from this list sorted by the expert provided ranks.
\end{enumerate}

It could be argued why we did not pick top 5 flags from the experts' list and chose to use the LLM. This is because:
\begin{enumerate}
    \item There are numerous ties in the expert-provided list and taking all tied values will significantly slow down the feature extraction.
    \item We used the LLM's ranking as a deterministic secondary signal to break ties. This provided a consistent and non-arbitrary method for resolving ambiguity instead of relying on random or alphabetical sorting.
\end{enumerate}

The next task was to generate relevant features from traces produced by enabling the flags. Without knowing the structure of trace data, it is difficult for an LLM to produce meaningful features. We could not find any documentation for the structure of these traces, and reverse engineering it from the V8 source code manually would be time-consuming. To address this, we utilized the ability of the LLMs to understand source code as follows.
\begin{enumerate}
    \item We provided the LLM with the V8 logging source code for each flag. Using Prompt \ref{prompt:flagprint}, we asked it to find and extract all print statements that emit trace information. This step automatically discovered the trace data's underlying schema. 
    \item Using this schema, we prompted the LLM again to propose features that aid in finding vulnerabilities (Prompt \ref{prompt:regexgen}).
\end{enumerate}
This feature list, consisting of regexes, was manually validated for accuracy against the discovered trace structure. This process yielded 49 dynamic features, out of which only 7 features required modification to conform to the trace structure. 

\subsection{Phase 4: Vectorization of Training/ Testing Data}
The feature vector $\mathbf{x}_{new}$ is generated by a feature extraction process, $\mathcal{F}$, that applies the static and dynamic feature regexes collected from Phase 2 and 3 to a given input, $I_{new}$. The feature extraction process is described below.
\begin{enumerate}
    \item \textbf{Static Feature Extraction:} Let $\mathcal{F}_{static}$ be the function implementing Algorithm \ref{alg:llm_feature_extraction}. It takes the source code of the input $I_{new}$ and returns the following static feature vector of dimension $n_s$.
    \[
    \begin{aligned}
    F_{\text{static}} &: I \to \mathbb{R}^{n_s}, \quad x_{\text{static}} = F_{\text{static}}(I_{\text{new}}) \\
    \end{aligned}
    \]
    \item \textbf{Dynamic Feature Extraction:} A second function, $\mathcal{F}_{dynamic}$ implementing Algorithm \ref{alg:llm_dynamic_feature_extraction}, executes the input $I_{new}$ with the selected V8 tracing flags, captures the resulting log output, and parses it to construct the dynamic feature vector of dimension $n_d$.
    \[
    \begin{aligned}
    F_{\text{dynamic}} &: I \to \mathbb{R}^{n_d}, \quad x_{\text{dynamic}} = F_{\text{dynamic}}(I_{\text{new}})\\
    \end{aligned}
    \]
\end{enumerate}

Therefore, we can represent the full feature extraction process for a new input $I_{new}$ as:

\[
F : I \to \mathbb{R}^{n_s + n_d}, \quad x_{\text{new}} = F(I_{\text{new}}) = [x_{\text{static}}, x_{\text{dynamic}}]
\]

\subsection{Phase 5: Building the Guidance Model}
In the state-of-the-art coverage guidance, the target is to execute as many unique edges as possible.  Mathematically, the objective is to find a new input, $I_{new}$, that expands the set of known code coverage from the corpus $\mathcal{C}$:
$$\textbf{Objective: } \text{Find } I_{new} \text{ such that } \text{Coverage}(I_{new}) \setminus \text{TotalCoverage}(\mathcal{C}) \neq \emptyset$$
The coverage guidance asks - \textbf{is this path new?}, where the proposed model asks - \textbf{does this code look dangerous?}. Hence, unlike the coverage guided signal, the guidance signal is now a vulnerability score computed by the following model. The fuzzer's new objective is to find inputs that maximize this score when applied to its feature vector $\mathbf{x}_{new}$:
$$\textbf{Objective: } \text{Find } I_{new} \text{ that maximizes } \mathcal{M}(\mathbf{x}_{new})$$

We implemented $M$ using the XGBoost Machine Learning model.

\begin{tcolorbox}[colback=gray!5!white, colframe=gray!50!black, title=RQ1: How to systematically derive a set of static and dynamic features most indicative of vulnerabilities in JS engine? ]
We answer this with a multi-phase, LLM-guided process. First, an iterative feedback loop with an LLM produced a list of 115 static code features from a corpus of 30 real-world V8 PoCs. Next, the LLM clustered those static features and used the clusters to recommended the five most relevant V8 runtime flags for tracing dynamic behavior. Finally, the LLM analyzed the V8 source code relevant to those flags to generate 49 regular expressions, which can be used to extract dynamic features from the trace logs. This process yielded a comprehensive set of static and dynamic features for our guidance model.
\end{tcolorbox}

\subsection{Feature-Guided Fuzzing}
We used the top selected features and the guidance model to develop a feature-guided fuzzer. This fuzzer was built on top of Fuzzilli, which has been utilized in multiple works in JS engine fuzzing for its demonstrated strength in generating syntactically and semantically valid inputs \cite{wachter2025dumpling, bernhard2022jit, wang2024optfuzz, wang2023fuzzjit}. Our fuzzing strategy is shown in Algorithm \ref{alg:fuzz}.

The fuzzing process is a exploration-exploitation loop where we dedicated 90\% of the effort to feature-guided exploitation and 10\% to coverage-based exploration. In this way, we allowed the features to mainly drive the fuzzer but prevented stagnation by occasionally introducing new code paths. Moreover, without occasional exploration, the feature-guidance may discard seeds which are predicted benign but some mutations away from becoming a high-risk one. The coverage guidance helps to keep these seeds alive as long as these explore new paths.

\begin{algorithm}[t!]
\caption{Hybrid Fuzzing with SHAP-Guided Feature Preservation}
\label{alg:fuzzing_loop_final}
\scriptsize
\begin{algorithmic}[1]
\State \textbf{Input:} Initial seed corpus $C_{seed}$
\While{time permits}
    \State $parent \leftarrow \text{SelectSeed}(C_{seed})$
    \State $F_{shap} \leftarrow \text{GetShapFeatures}(parent)$ \Comment{Get important feature names for parent}
    \State $mutant \leftarrow \text{Mutate}(parent)$
    
    \If{\text{ExecutesAndCrashes}(mutant)}
        \State Add $mutant$ to corpus and \textbf{continue}
    \EndIf
    
    \If{$\text{rand}() \le 0.1$} \Comment{10\% exploration}
        \If{\text{HasNewCoverage}(mutant)}
            \State Add $mutant$ to corpus
        \EndIf
    \Else \Comment{90\% exploitation}
        \If{static features of $mutant$ in $F_{shap}$ are within 1$\sigma$ of parent's} \Comment{\textit{Faster filter}}
            \If{$\text{PredictScore}(mutant) \ge \text{GetScore}(parent)$}  \Comment{\textit{Slower filter}}
                 \If{dynamic features of $mutant$ in $F_{shap}$ are within 1$\sigma$ of parent's}
                    \State Add $mutant$ to corpus
                 \EndIf
            \EndIf
        \EndIf
    \EndIf
\EndWhile
\end{algorithmic}
\label{alg:fuzz}
\end{algorithm}

In this work, to prevent mutations from destroying high-risk features as discussed in \S \ref{motivation}, we extended the concept of aspect-preserving mutation proposed by Park et al. \cite{park2020fuzzing} to feature-preserving mutation. Unlike aspects, features are more difficult to preserve because 1) static features often do not map to a single code element but may span multiple parts of a seed 2) dynamic features are hard to control and associate with corresponding code changes. Hence, instead of pro-actively enforce feature-preservation, we adopted a filtering-based approach. After each mutation, each mutated unique seed is checked against its parent for feature preservation and those failing a threshold are discarded. 

Since running this check on every seed can be computationally expensive, we implemented a two-stage filter. Dynamic analysis is computationally expensive due to the overhead of code execution and tracing , whereas static analysis is fast. Our filter leverages this difference:

\begin{itemize}
    \item Fast Static Filter: A mutation first undergoes a static analysis. Then, a check ensures its static features remain within a one-standard-deviation ($1\sigma$) tolerance of the parent's values. We empirically selected this threshold after observing that looser thresholds (e.g.$2\sigma$, $3\sigma$, etc.) accept nearby everything while tighter thresholds are too restrictive. However, we recommend this parameter as tunable.
    \item Slow Dynamic Filter: If a mutation passes the static check, it undergoes a full dynamic analysis adhering to the $1\sigma$ tolerance. We further verify that its predicted vulnerability score has not decreased.
\end{itemize}
        
Also, all the features do not go through the preservation check. This is because 1) preserving all the features can be overly restrictive, 2) while our initial analysis identified a globally important minimal feature set, all of these features are not critical for every vulnerability. Hence, we used SHapley Additive exPlanations (SHAP) to provide instance-specific guidance \cite{lundberg2017unified,zhang2023shapfuzz}. For each parent seed, we identified the feature subset that SHAP deemed most responsible for its high vulnerability score, preserving 90\% of the vulnerability score coverage. The subsequent preservation checks focused only on this subset.

Only mutants that pass this multi-stage process are added to the seed corpus. This ensures that the fuzzer prioritizes only high-risk seeds.

\section{Evaluation}
In this section, We show that a fraction of the features (25\%) collected using only 5 runtime flags is sufficient to achieve excellent predictive performance and faster fuzzing results. We answer the following questions that lead to this conclusion. 
\begin{enumerate}
    \item[RQ2] How effective is the guidance model in terms of predictive performance?
    \item[RQ3] Do both static and dynamic features contribute to its performance?
    \item[RQ4] What is the minimal feature set beyond which additional features yield diminishing returns?
    \item[RQ5] Can the feature-guided fuzzing process find real bugs?
    \item[RQ6] What is the runtime overhead of collecting these features?
\end{enumerate}


\begin{table}[t]
\centering
\caption{Dataset distribution and experimental setup. Numbers indicate unique proof-of-concept (PoC) programs; positives are crash-inducing, negatives are non-crashing. \% for positives indicate proportion relative to all samples in that fold.}
\label{tab:dataset_distribution}
\resizebox{\linewidth}{!}{%
\begin{tabular}{lccccc}
\toprule
\textbf{Metric} & \textbf{Fold 1} & \textbf{Fold 2} & \textbf{Fold 3} & \textbf{Fold 4} & \textbf{Total} \\
\midrule
Positive Samples (count, \%) & 29 (2.04\%) & 46 (1.82\%) & 48 (2.04\%) & 37 (1.48\%) & 160 \\
Negative Samples & 2,064 & 2,211 & 2,159 & 2,064 & 8,498 \\
Class Ratio (Pos / Neg) & 1:71 & 1:48 & 1:45 & 1:56 & 1:53 \\
JS Engine Versions Tested & 7.4.46--11.0.140 & 11.1.92--11.8.149 & 11.9.166--12.7.173 & 12.8.168--13.6.33 & --- \\
Fold Time Range & $\leq$ Nov~2022 & Dec~2022--Aug~2023 & Sep~2023--May~2024 & June~2024--Mar~2025 & --- \\
\bottomrule
\end{tabular}%
}
\end{table}

\subsection{Experimental Setup}
The time-controlled experiments were run on an Intel Core i7-1065G7 processor (4 cores) with 16 GB of memory, running Ubuntu 22.04. For the fuzzing experiment, we used an Ubuntu 22.04 machine with a 16 core processor and 128 GB memory. The datasets, methods compared, metrics, and statistical methods are discussed below. 

\subsubsection{Dataset}
For our experiments, the data is divided into four folds for cross-validation with a walk-forward protocol \cite{falessi2020need}. Table \ref{tab:dataset_distribution} shows the distribution of the datasets in each folds. As detailed in the table, each fold contains dataset with PoCs and JS engine test cases in the corresponding time range. The dataset at the $i-1^{th}$ time range is used for training, and the $i^{th}$ time range is used for testing. This chronological split ensures that the model is always validated on data collected after the training period. The temporal width of folds 2-4 is about 9 months. Fold 1 expands over 3 years due to the lack of positive examples in this time range and to provide sufficient initial training data to the predictor. Notably, the dataset is highly imbalanced with a positive to negative ratio of at least $1:48$. This creates a realistic ground for evaluation that is equivalent to rarity of crashes seen in fuzzing sessions.

\subsubsection{Machine Learning Algorithms}
For the predictive evaluation, we use the XGBoost machine learning algorithm, which is an ensemble learning approach built on boosting \cite{ChenG16}. This algorithm sequentially combines decision trees applying gradient boosting, where each subsequent tree is trained to reduce the residual error of the previous tree. We implemented XGBoost using Scikit-learn \footnote{\url{https://xgboost.readthedocs.io/en/stable/python/sklearn_estimator.html}}. XGBoost offers several benefits for our evaluation, which are as follows.  

\begin{enumerate}
    \item \textbf{Handling of Class Imbalance:}  Scikit-learn's XGBoost includes the $scale\allowbreak\_pos\allowbreak\_weight$ hyperparameter, which allows us to increase the penalty for misclassifying the rare positive class. 
    \item \textbf{High-predictive Performance:} XGBoost is widely regarded as the state-of-the-art model for tabular data. Shwartz-Ziv and Armon showed that XGBoost can be as effective as deep learners for tabular data, in terms of both predictive performance and training time \cite{shwartz2022tabular}.
    \item \textbf{Model Interpretability:} We use the XGBoost feature importance score to answer RQ4. This allows us to run studies comparing different fractions of top features and compare their performances.
\end{enumerate}

\subsubsection{Metrics}
We use precision, recall, and false alarm rate to evaluate performance, which are defined below. 

\begin{itemize}
    \item Precision: It is the ratio of correctly predicted positive observations (True Positive $TP$) to the total predicted positive ones (True Positive $TP$+ False Positive $FP$). It answers the question: ``of all the cases the model predicted as positive, how many were actually positive?''.
    \begin{equation}
    \text{Precision} = \frac{TP}{TP + FP}
    \end{equation}
    \item \textbf{Recall:} Recall is sometimes also referred to as sensitivity or true positive rate which is defined as follows. Recall assesses the model's ability in detecting positive cases.

    \begin{equation}
    \text{Recall} = \frac{TP}{TP + FN}
    \end{equation}

    \item \textbf{False Alarm Rate:} This is the false positive rate, which measures the percent of actual negatives that were classified as positive. This gives the likelihood of predicting a positive value incorrectly.
    \begin{equation}
    \text{False Alarm Rate} = \frac{FP}{FP + TN}
    \end{equation}
    
\end{itemize}
\subsubsection{Statistical Analysis}
We statistically compared our methods using the Wilcoxon Rank-Sum test \cite{wilcoxon1945individual, mann1947test}. We selected this test because it is non-parametric, which means that it makes no assumption about the distribution of the metrics compared. The null hypothesis of this test is -  there is no difference between the tested and the baseline. Wilcoxon Rank-Sum test works by ranking all data points from both groups together from smallest to largest. The sum of the ranks for each group is then compared, and if one group's ranks are consistently higher, it suggests a statistically significant difference between the two groups. 

The test results in a p-value. If the p-value is below a significance level, the null hypothesis can be rejected. As commonly used, we choose the significance level as 0.05 \cite{conover1999practical}.  

\subsection{Results}
\subsubsection{[RQ2] How effective is the guidance model in terms of predictive performance?}
\begin{figure}[H]
    \centering
    \includegraphics[width=0.8\linewidth]{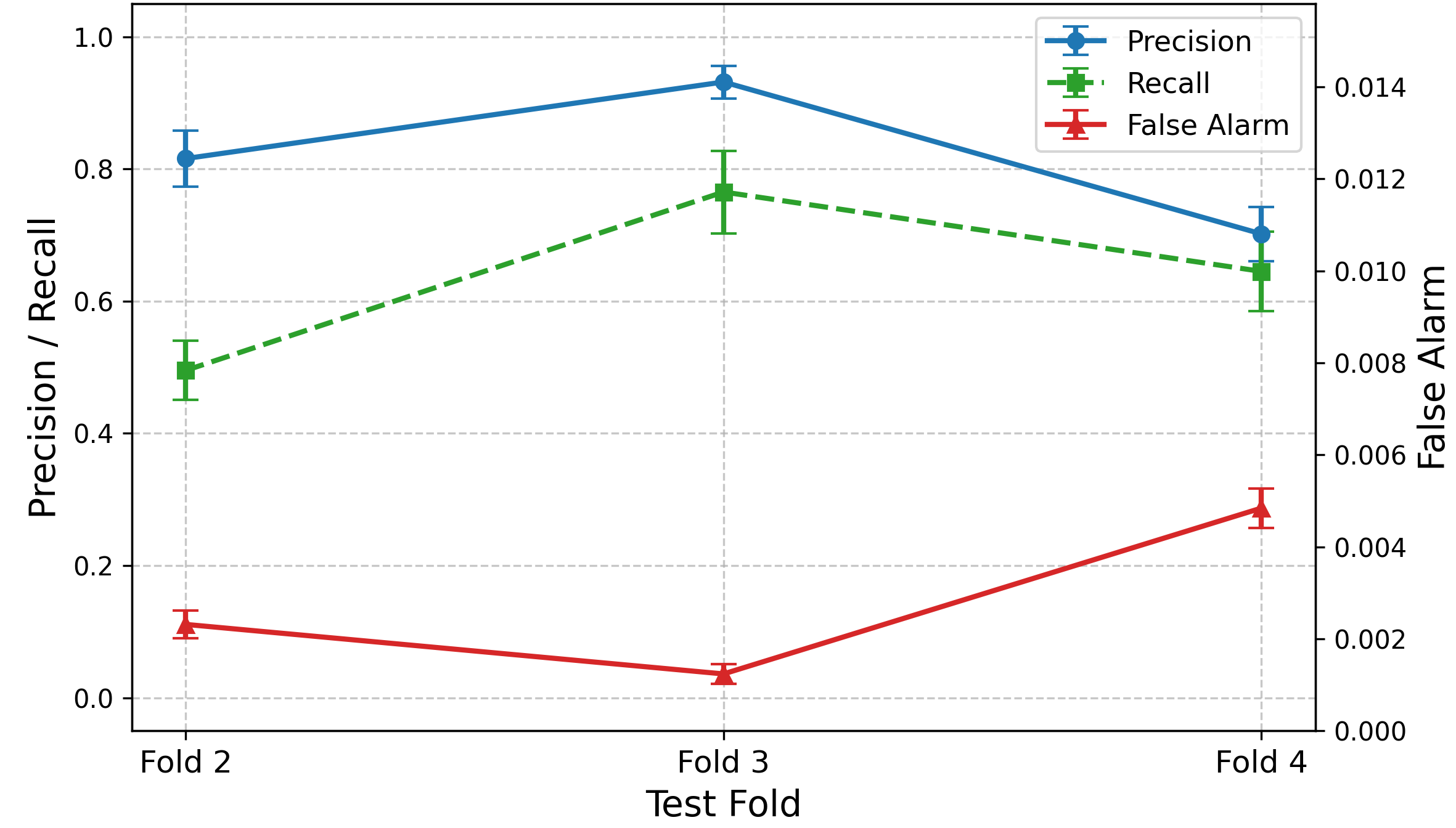}
    \caption{Precision, recall, and false alarm rates over time-aware cross-validation folds. Precision remains consistently high, recall shows moderate variation, and false alarm rates are low, indicating strong predictive reliability of the guidance model.}
    \label{fig:metrics_combined_top_fraction}
\end{figure}

Figure \ref{fig:metrics_combined_top_fraction} shows the precision, recall, and false alarm across all the folds. The guidance model shows consistently high precision ($\approx 0.74–0.94$) across all folds. The high precision indicates that if the model makes a prediction, it is more likely to be correct. The false alarm rate is also very low ($\le 0.017$).

Recall is slightly lower ($\approx 0.52–0.81$). Upon exploring the PoCs the model classified incorrectly, we identified the following potential reasons.

\begin{enumerate}
    \item Limited PoC diversity: features were extracted from a small sample of 30 PoCs, which may not fully represent every types of vulnerabilities.
    \item Concept Drift: The time-aware folds introduce future samples with potentially different patterns not seen during training. This indicates that recall could potentially be improved by online learning techniques such as incremental learning. We leave this for future work (See \S \ref{futurework}).
    \item Data Imbalance: In the first fold, data imbalance is the highest, which contributes to the low recall.
\end{enumerate}

Overall, the model is cautious in detecting a vulnerability triggering inputs. During fuzzing, it means reduced wasted effort in seeds that have less potential to trigger a vulnerability \cite{bohme2017directed}.  

\begin{tcolorbox}[colback=gray!5!white, colframe=gray!50!black, title=RQ2: How effective is the guidance model in terms of predictive performance?]
While the model cannot find all the vulnerabilities, it is precise. Hence, it potentially reduces the time wasted on non-triggering inputs. Improving PoC diversity and retraining with temporally broader data could improve recall without losing too much precision.
\end{tcolorbox}

\begin{table*}[ht]
\centering
\caption{Mean and standard deviation (±) of Precision (Prec.), Recall (Rec.), and False Alarm rate (FA) across three folds, reported as decimals between 0 and 1 rounded to two decimal places. The values in parentheses represent the standard deviation of the last two digits. The combined feature method achieves the highest precision and recall with minimal false alarms.}
\label{tab:summary_compact}
\setlength{\tabcolsep}{2pt}
\scriptsize
\begin{tabularx}{\textwidth}{@{} l @{\hspace{4pt}} c *{3}{S[table-format=1.2, table-auto-round, table-column-width=1cm]} *{3}{S[table-format=1.2, table-auto-round, table-column-width=1cm]} *{3}{S[table-format=1.2, table-auto-round, table-column-width=1cm]} @{}}
\toprule
\multirow{2}{*}{Method} & \multirow{2}{*}{\#Feat.} & \multicolumn{3}{c}{Fold 2} & \multicolumn{3}{c}{Fold 3} & \multicolumn{3}{c}{Fold 4} \\
\cmidrule(lr){3-5} \cmidrule(lr){6-8} \cmidrule(lr){9-11}
 & & {Prec.} & {Rec.} & {FA} & {Prec.} & {Rec.} & {FA} & {Prec.} & {Rec.} & {FA} \\
\midrule
static & 115 & 0.29 \pm 0.04 & 0.28 \pm 0.06 & 0.02 \pm 0.00 & 0.72 \pm 0.18 & 0.71 \pm 0.04 & 0.01 \pm 0.01 & 0.26 \pm 0.07 & 0.50 \pm 0.10 & 0.03 \pm 0.01 \\
dynamic & 49 & 0.26 \pm 0.02 & 0.47 \pm 0.03 & 0.03 \pm 0.00 & 0.43 \pm 0.02 & 0.71 \pm 0.03 & 0.02 \pm 0.00 & 0.17 \pm 0.02 & 0.74 \pm 0.05 & 0.07 \pm 0.01 \\
combined & 164 & 0.89 \pm 0.05 & 0.52 \pm 0.04 & 0.00 \pm 0.00 & 0.94 \pm 0.02 & 0.81 \pm 0.03 & 0.00 \pm 0.00 & 0.74 \pm 0.05 & 0.70 \pm 0.07 & 0.01 \pm 0.00 \\
random & 0 & 0.02 \pm 0.00 & 0.53 \pm 0.09 & 0.50 \pm 0.01 & 0.02 \pm 0.00 & 0.45 \pm 0.06 & 0.50 \pm 0.01 & 0.02 \pm 0.00 & 0.50 \pm 0.10 & 0.49 \pm 0.01 \\
random\_features & 164 & 0.02 \pm 0.02 & 0.01 \pm 0.02 & 0.02 \pm 0.00 & 0.04 \pm 0.02 & 0.03 \pm 0.02 & 0.02 \pm 0.00 & 0.03 \pm 0.04 & 0.04 \pm 0.05 & 0.02 \pm 0.01 \\
\bottomrule
\end{tabularx}
\end{table*}

\begin{table*}[ht]
\centering
\caption{Wilcoxon rank-sum test $p$-values comparing Precision (Prec.), Recall (Rec.), and False Alarm (FA) between methods. Significant results ($p<0.05$) are shown in \textbf{\textcolor{green!60!black}{bold green}}, non-significant ($p \geq 0.05$) in \textcolor{red}{red}.}
\label{tab:wilcoxon_results}
\scriptsize
\setlength{\tabcolsep}{4pt}
\begin{tabularx}{\textwidth}{l l *{3}{S[table-format=1.2e-1, table-align-text-post=false]}}
\toprule
Comparison & Method & {Prec.} & {Rec.} & {FA} \\
\midrule
\multirow{4}{*}{vs. Static} 
 & Dynamic          & {\textbf{\textcolor{green!60!black}{$<1e-4$}}} & {\textbf{\textcolor{green!60!black}{$<1e-4$}}} & {\textbf{\textcolor{green!60!black}{$<1e-4$}}} \\
 & Combined         & {\textbf{\textcolor{green!60!black}{$<1e-4$}}} & {\textbf{\textcolor{green!60!black}{$<1e-4$}}} & {\textbf{\textcolor{green!60!black}{$<1e-4$}}} \\
 & Random           & {\textbf{\textcolor{green!60!black}{$<1e-4$}}} & {\textcolor{red}{0.81}} & {\textbf{\textcolor{green!60!black}{$<1e-4$}}} \\
 & Random\_features & {\textbf{\textcolor{green!60!black}{$<1e-4$}}} & {\textbf{\textcolor{green!60!black}{$<1e-4$}}} & {\textcolor{red}{0.27}} \\
\midrule
\multirow{4}{*}{vs. Dynamic} 
 & Static           & {\textbf{\textcolor{green!60!black}{$<1e-4$}}} & {\textbf{\textcolor{green!60!black}{$<1e-4$}}} & {\textbf{\textcolor{green!60!black}{$<1e-4$}}} \\
 & Combined         & {\textbf{\textcolor{green!60!black}{$<1e-4$}}} & {\textbf{\textcolor{green!60!black}{$3.04e-2$}}} & {\textbf{\textcolor{green!60!black}{$<1e-4$}}} \\
 & Random           & {\textbf{\textcolor{green!60!black}{$<1e-4$}}} & {\textbf{\textcolor{green!60!black}{$4.00e-4$}}} & {\textbf{\textcolor{green!60!black}{$<1e-4$}}} \\
 & Random\_features & {\textbf{\textcolor{green!60!black}{$<1e-4$}}} & {\textbf{\textcolor{green!60!black}{$<1e-4$}}} & {\textbf{\textcolor{green!60!black}{$<1e-4$}}} \\
\bottomrule
\end{tabularx}
\end{table*}
\vspace{-0.4cm}

\subsubsection{[RQ3] Does both static and dynamic features contribute to its performance?}
We performed an ablation study comparing static, dynamic, and the combined model. We used two baselines, which are \textit{random} and \textit{random\_features}. The \textit{random} baseline, which classifies PoCs randomly, validates if the predictive performance is due to chance. The model must perform better than the random baseline to be meaningful. The \textit{random\_features} baseline generates random predictions but adjusts the positive prediction threshold based on the average positive probability predicted by a model trained on the features. This baseline reflects both the data’s class imbalance and some feature-informed signal. It tests if the model learns useful signals from the LLM-generated features, rather than reflecting the data’s class distribution. 

Table \ref{tab:summary_compact} shows the ablation study results. In most of the cases, the static, dynamic, and combined models outperform the random and random features baseline, indicating the meaningfulness of the feature set and predictive model. We can observe that combining both static and dynamic significantly improves performance. The combination approximately doubles the precision in almost all cases. In fold 2 and 3, it also increases the recall by a significant amount. In fold 4, while the recall is slightly reduced, the combined model remains the best in terms of precision, recall, and false alarm. Overall, we conclude that both static and dynamic models contribute to the performance. Furthermore, the model with static features has slightly higher precision, and the dynamic model usually has higher recall.

\begin{tcolorbox}[colback=gray!5!white, colframe=gray!50!black, title=RQ3: Do both static and dynamic features contribute to its performance?]
Both static and dynamic features contribute to the performance as the combined model significantly outperforms both. 
\end{tcolorbox}

\begin{figure}[t!]
    \centering
    \begin{subfigure}[b]{0.48\textwidth}
        \centering
        \includegraphics[width=\textwidth]{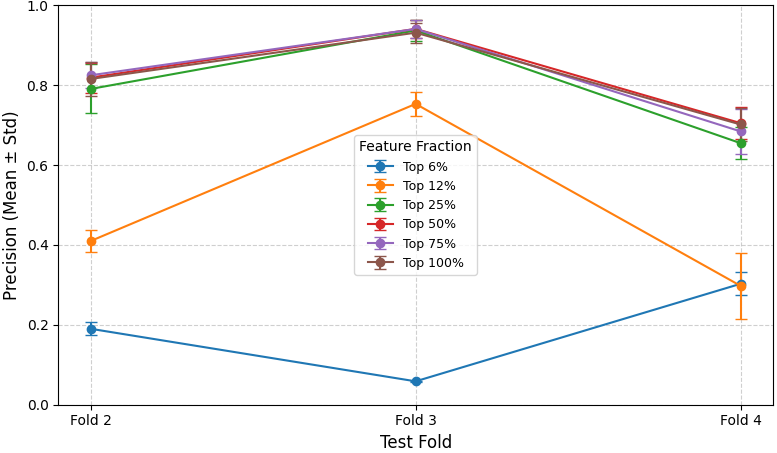}
        \caption{Precision vs. fold.}
        \label{fig:precision-features}
    \end{subfigure}
    \hfill
    \begin{subfigure}[b]{0.48\textwidth}
        \centering
        \includegraphics[width=\textwidth]{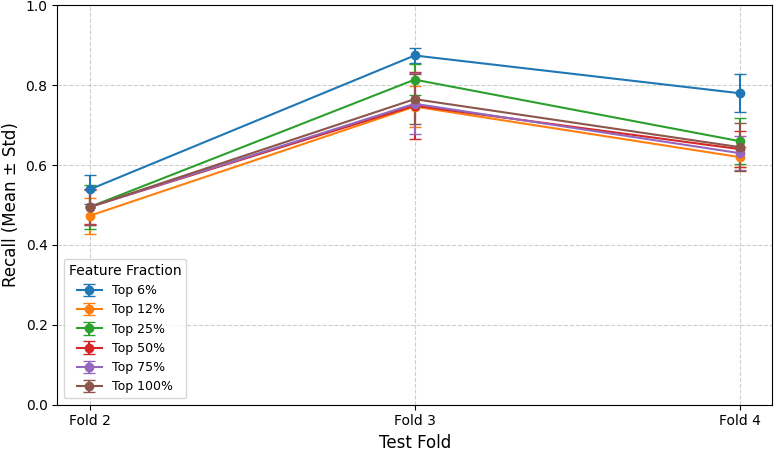}
        \caption{Recall vs. fold.}
        \label{fig:recall-features}
    \end{subfigure}
    
    \begin{subfigure}[b]{0.5\textwidth}
        \centering
        \includegraphics[width=\textwidth]{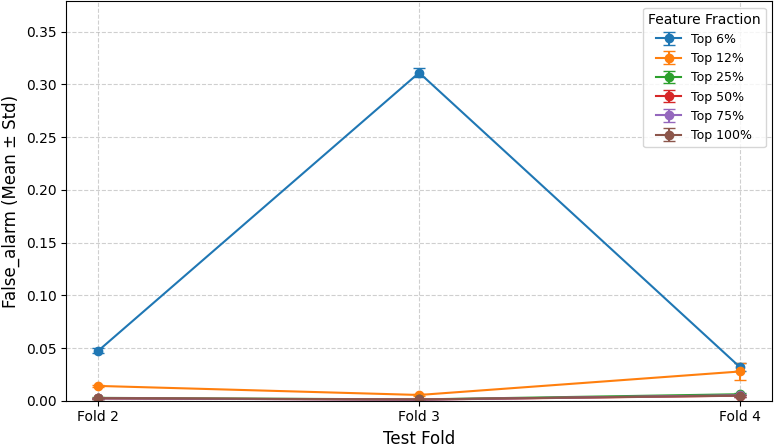}
        \caption{False alarm rate vs. fold.}
        \label{fig:false-alarm-features}
    \end{subfigure}
    
    \caption{Comparative evaluation of model performance across three test folds for different top-ranked feature fractions. 
    Precision, recall, and false alarm plateau for Top 25\%–100\% of features, with nearly overlapping means and low variance, indicating diminishing returns from including the remaining 75\% of features.}
    \label{fig:all-feature-fractions}
    \vspace{0.1cm}
\end{figure}

\begin{figure}[H]
    \centering
    \includegraphics[width=\textwidth]{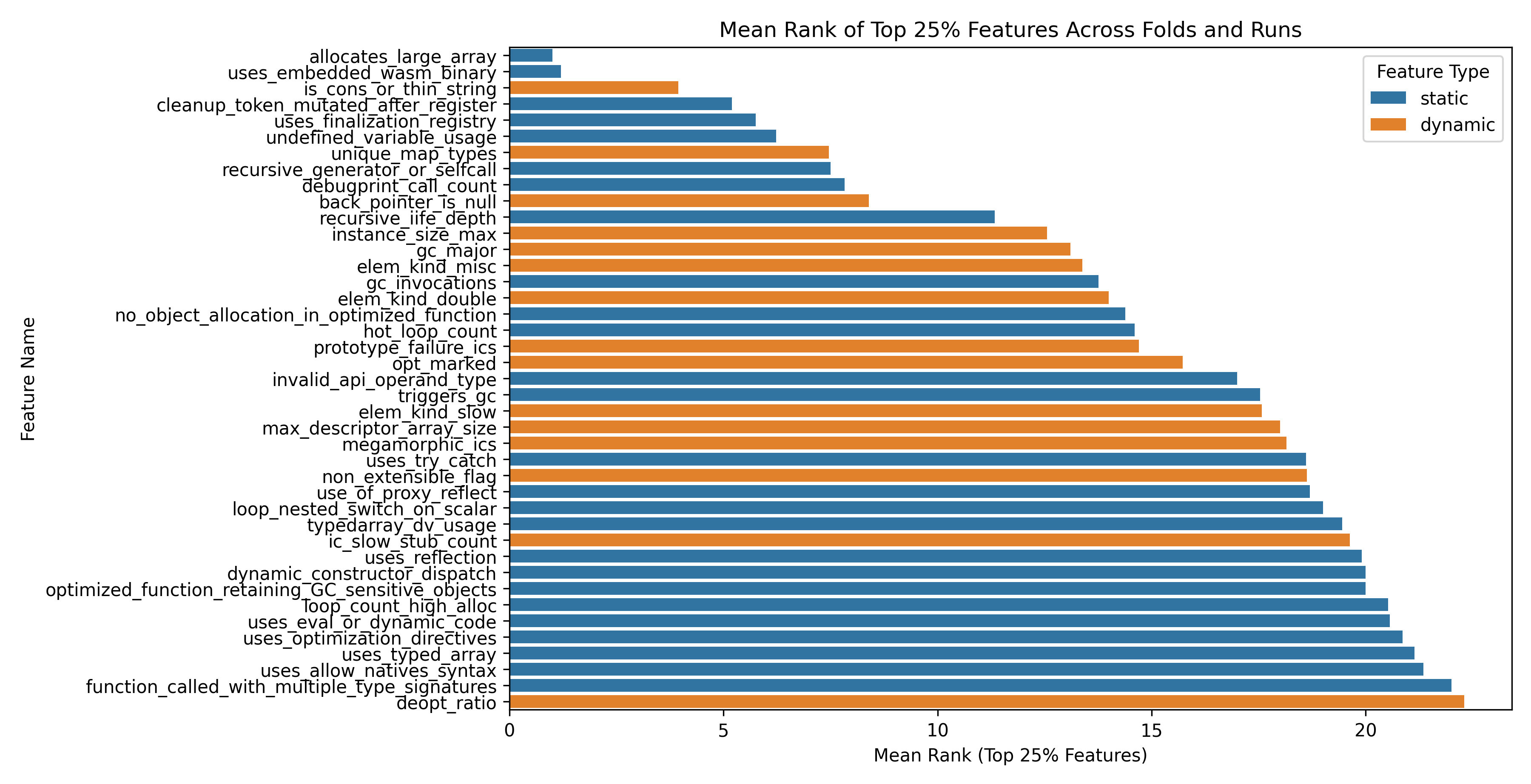}
    \caption{Mean rank of the top 25\% features across different folds and runs. Features are categorized as either static or dynamic, providing insight into their relative importance for vulnerability prediction.}
    \label{fig:feature-ranking}
\end{figure}
\subsubsection{[RQ4] What is the minimal feature set beyond which additional features yield diminishing returns?}
To answer this question, we compare the metrics across all the folds for varying feature fractions. To do this, XGBoost model is applied on a fold and all the $N$ features are ordered by the average feature importance score (information gain). Then, we run experiment for each fold for each $N\times\%x$ features, where $x$ is the percent of features to test. We start with 6\% of the features and double it as: [6\%, 12\%, 25\%, 50\%, 75\%, 100\%].

Figure \ref{fig:all-feature-fractions} shows the results. It is evident that using top 25\% features provides with the most optimal balance of precision, recall, and false alarm, along with the number of features. Adding features beyond this has diminishing returns because the performance improves minimally with twice the number of features, which has significant runtime overhead. 

For lower number of features, there is large performance degradation in at least one of the metrics. For example, top 6\% performs poorly in terms of  precision and false alarm, leading to a large waste of resource on non-triggering inputs. Similarly, top 12\% results in low precision. 

Figure \ref{fig:feature-ranking} shows the top 25\% features, extracted and ordered by their mean rank across all folds. Out of 164 features, only 41 features are selected, which are a combination of static and dynamic features. 15 out of 41 features are dynamic, eliminating $\approx 69\%$ of the dynamic features. Additionally, $\approx 77\%$ of the static features can be removed. Thus, by eliminating a large number of features, the efficiency of the training model and predictions are improved, which in terms can improve the fuzzing throughput. Moreover, we see that about 31\% dynamic features are retained compared to 23\% static features. This indicates the proportional importance of our dynamic features compared to static ones because a randomly selected dynamic feature is about 35\% more likely to be important than a randomly selected static one.

\begin{tcolorbox}[colback=gray!5!white, colframe=gray!50!black, title=RQ4: What is the minimal feature set beyond which additional features yield diminishing returns?]
About 75\% of the features can be discarded without significantly impacting predictive performance. With limited resources, prioritizing dynamic feature collection offers more benefits because they have a moderately higher retention rate in the top features set.
\end{tcolorbox}

\subsubsection{RQ5: Can the feature-guided fuzzing process find real bugs?}
\begin{table}[h]
\centering
\caption{Bug-finding results over 96 hours.}
\begin{tabular}{l c c}
\hline
\textbf{Fuzzer} & \textbf{Crashes Found} & \textbf{Time (h)} \\
\hline
Our Fuzzer     & 2 & 5.0, 23.5 \\
Fuzzilli       & 1 & 94.0 \\
FuzzJIT        & 0 & -- \\
DIE            & 1 & 87.0 \\
\hline
\end{tabular}
\label{tab:fuzzres}
\end{table}
We performed a prudence check by running our feature-guided fuzzer. Similar to Wang et al. \cite{wang2024optfuzz}, we observed that running the state-of-the-art fuzzers for a week on the latest V8 did not yield any crashes. Hence, for a fair comparison with the baselines, we opted to use an older (2023) version of V8. In this setup, the seed corpus, selected feature sets, training data, predictor, and baselines were all derived from a version earlier than the one under test, avoiding any leakage of information. The results are shown in Table \ref{tab:fuzzres}. Our fuzzer was able to find 2 crashes within 24 hours, where the first crash was found within $\approx 5$ hours. Fuzzilli and DIE both found one crash in $\approx 94$ and $\approx 87$ hours, respectively. In our experiments, FuzzJIT was not able to find any crashes within the 96 hour fuzzing campaign. We manually inspected the crashes and found that these were semantically equivalent variants of already reported crashes. Therefore, we did not report them. 

To further evaluate bug finding ability, we ran our fuzzer on a latest version of V8 for 24 hours. It discovered a crash (\textit{Issue ID: 450017974}) within 5 hours. The crash was reported to the V8 team where it was found to be a valid crash with top priority and severity ($P1$ and $S1$). Further analysis of the crashing seed showed that our vulnerability predictor assigned high probability to it, indicating that the crash was discovered by our proposed mechanism rather than by random mutations.

Hence, our fuzzer can 1) find real bugs in the wild, and 2) find crashes faster as seen in the limited time runs in our experiments. 

\begin{tcolorbox}[colback=gray!5!white, colframe=gray!50!black, title=RQ5: Can the feature-guided fuzzing process find real bugs?]
Feature-guided fuzzing finds real crashes faster.
\end{tcolorbox}
\begin{table}[b!]
\caption{Execution rate comparison. Higher is better.}
\label{tab:exec-rate}
\centering
\begin{tabular}{lcc}
\hline
Setup                   & Exec. Rate & Change vs Baseline \\
\hline
Baseline (No Features)  & 175        & --                 \\
All Features (164)      & 111.97      & 36\% lower       \\
Top 25\% Features (41)  & 187        & 6.9\% higher       \\
\hline
\end{tabular}

\end{table}
\subsubsection{RQ6: What is the runtime overhead of collecting these features?}
To measure the runtime impact of feature collection during fuzzing, we compared the execution rate of our approach with a coverage-guided baseline (Fuzzilli) with no features. We measured the execution rate of both the full feature and top feature sets and compared with the baseline. Table \ref{tab:exec-rate} shows the results. Interestingly, while using the full feature set reduces the execution rate by 36\%, the top 25\% features achieves a slightly higher execution rate than the baseline. This is because the coverage-instrumented engine, which is a primary source of significant slowdown \cite{wang2021riff}, can mostly be avoided in feature-guided executions. Using the top 25\% is also about 1.56 times faster than using all the features. Therefore, we conclude that incorporating selected features into the fuzzing process does not incur significant runtime overhead and can even improve execution efficiency.
\begin{tcolorbox}[colback=gray!5!white, colframe=gray!50!black, title=RQ6: What is the runtime overhead of collecting these features?]
Computing the selected features during fuzzing has negligible runtime overhead and can improve the overall execution rate.
\end{tcolorbox}

\section{Discussion}

\subsection{Threats to Validity}
In this section, we address the potential limitations of our study and the way we addressed them. Any conclusions made from this study should take into account the following issues.

\subsubsection{Internal Validity}
We identify two potential sources of bias in the generated features. Firstly, the features were discovered from a set of 30 PoCs and a different sample could have led to a different feature set. While we do not claim that our features are the most optimal ones, these produce strong precision, recall, and false alarm in an unseen dataset. As our initial 30 PoCs were randomly selected from an exhaustively collected set, obtaining an effective set of features just by chance is likely mitigated. Secondly, manual curation to remove overfitting features depends on researcher subjectivity. While it was not possible to completely eliminate this bias, most of the overfitting removal was done in an objective manner, for example, removing variable names, uncommon features used in only one PoC etc. For the remaining removals, we consulted the industry experts, co-author and the LLM itself. 

Another source of bias is the confounding information in the LLM prompts we used. For example, since the written bug reports are used with the PoC source code for static feature generation, the LLM may learn the features from the bug reports instead of the source code. We mitigate this in the iterative feedback loop, where we provide the LLM with the feature set only and drive the LLM to use the bug report only as a context for feature generation. 

\subsubsection{External Validity}
Our dataset and analysis are limited to the V8 JS engine, which holds about 75\% market share for web browsers, expanding across Google Chrome, Chromium, Edge, and Opera. It is also used in server-side JS environment, PDF readers, desktop and mobile application. While this ubiquity makes V8 an impactful choice, the resulting features, particularly the dynamic ones derived from the V8 flags, may not transfer directly to other JS engines like SpiderMonkey or JavaScriptCore. However, the overall data-centric methodology is applicable, and specific feature set would likely need to be regenerated. We leave the application of our methodology to other engines for future work.   

Our feature generation pipeline is dependent on GPT-4.1. This poses a generalizability risk that a different LLM may produce a different but ineffective set of features. We mitigate this risk by iterative validation on unseen PoCs and manual curation, a strategy shown to improve LLM generalizability in other contexts \cite{xu2024enhancing, wu2025step}. 

A prediction model trained on historical data may see its performance degrade as new classes of bugs emerge. We explicitly address this by using a time-aware, walk-forward cross-validation technique. Our experiments suggest that by re-training every nine months, the model is able to keep up with the newly introduced bugs.

\subsubsection{Construct Validity}
Our claim is that a set of features can be considered as a proxy for the "dangerousness" of a seed. Our set of features serves as a proxy for this abstract concept. It is possible that this proxy is incomplete and misses some vulnerability patterns. However, its $\approx 85\%$ precision and $<1\%$  false alarm shows that, while not exhaustive, it captures the majority of high-risk seeds to guide fuzzing.

Another threat to construct validity is dataset representativeness. We use the mjsunit regression suite and the publicly reported PoCs as the primary sources of negative and positive samples. We acknowledge that this sampling cannot capture all possible vulnerabilities (e.g. undisclosed exploits) or the full characteristics of benign JS in the wild. However, our positive samples were collected from the Chromium bug repository, reproduced systematically, excluding only irreproducible PoCs. The mjsunit test suit is widely used as seed corpus because it contains both simple and adversarial code \cite{lee2020montage, wen2023evaluating}. In other words, it contains hard negatives that are difficult for the predictive model. Prior studies support this, for example, Montage shows that most buggy fragments come from the JS engine tests, and DIE's semantic preservation infers seed corpus with regression tests exhibit high-risk behavior. Hence, using this makes our findings more meaningful compared to using trivial and easy-to-predict negative sample. 

\subsection{Future Work}
\label{futurework}
\subsubsection{Online Learning:} Our approach addresses concept drift by re-training the model periodically (e.g. every nine months in our experiments). A more dynamic approach would be to consider incremental learning techniques. Prior research has shown that active learning can effectively address the concept drift issue in areas such as malware detection \cite{chen2023continuous} and defect prediction \cite{xu2018cross, mei2024cross}.
\subsubsection{Explanation:} The features used in this study could be extended and used to explain the root cause of a given crash. The matrix of Shapley values we get from SHAP can help to identify the specific features that contribute most to a high vulnerability score of a crashing input. These top contributing features can be fed back into an LLM to generate natural language hypothesis about the potential root cause. 

\subsubsection{Crash Triaging:} Once a crash occurs, it is time-consuming to determine if the crash corresponds to a vulnerability \cite{park2019empirical}. As our predictive model is trained on vulnerable PoCs, an important future direction is to evaluate its effectiveness in predicting whether a crash indicates a vulnerability.

\subsubsection{Generalization to Other JS Engines and Domains:} While the current feature set is derived using V8, cross-engine feature transfer could be explored to assess how well these features transfer to other JS engines. This could help to identify universal indicators of JS vulnerabilities versus the engine-specific ones. Moreover, since the LLM-guided feature extraction process is not limited to JS, it could be adapted for fuzzing other software with large search spaces such as operating system kernels, web browsers, or database systems. 
\section{Conclusion}
This paper addresses the inefficiencies involved with fuzzing JS engines which involves the failure to prioritize high-risk paths in traditional coverage guided and limited mitigations from manual rules and observations in the literature. We conjecture that such methods spend a lot of effort mutating irrelevant codes, thereby wasting resources. To overcome this, we introduced a novel data-centric paradigm, shifting the focus from coverage to causes by asking not ``is this path new?'' but rather ``does this code look dangerous?''

Our contribution is a systematic, LLM-guided methodology to extract a compact set of static and dynamic features from a historical corpus of real-world vulnerabilities. By training an XGBoost classifier on these features, we developed a high-fidelity guidance model that predicts the likelihood of an input triggering a crash. Our evaluation, conducted using a time-aware, walk-forward protocol that mimics real-world scenarios, demonstrates the effectiveness of this approach.

Our findings show that:
\begin{itemize}
    \item The guidance model is highly effective with over 85\% precision with a false alarm rate below 1\%.
    \item A hybrid approach is essential, as the combination of static and dynamic features significantly outperforms models built on either feature type alone.
    \item There are diminishing returns to adding features. A minimal set of the top 25\% of features provides a near-optimal balance of predictive power and efficiency.
    \item This feature-guided approach is practical and efficient. We found bugs both in an old and a recent version of V8 within hours instead of weeks or months.
\end{itemize}

In conclusion, we recommend that future fuzzing approaches prioritize cause over coverage. This is a fundamental shift from heuristic-driven exploration to adaptive, data-driven fuzzers that can keep up with ever-growing software complexity. This research demonstrates that the future of fuzzing lies not in exploring more, but in exploring less but smarter by leveraging historical data as a predictive guide to efficiently discover vulnerabilities. 
\newpage






{\footnotesize
\bibliographystyle{unsrtnat}
\bibliography{references}
}
\end{document}